\RequirePackage{lineno}\newdimen\linenumbersep\linenumbersep=2pt
\documentclass[preprint,4p,number,sort&compress,longtitle,times]{elsarticle}

\usepackage{comment}
\usepackage{subfigure}
\usepackage{mathtools}
\usepackage{tikz}
\usepackage[compat=1.1.0]{tikz-feynman}
\usetikzlibrary{shapes,arrows,positioning,automata,backgrounds,calc,er,patterns}

\usepackage{array, booktabs, color, eurosym, graphicx, longtable, pbox, pdfpages, siunitx, url, verbatim}

\usepackage[colorlinks=true,citecolor=blue,filecolor=blue,linkcolor=blue,urlcolor=blue,pdftex]{hyperref}
\usepackage{xspace}

\newcommand{\cevns}{\protect{CE$\nu$NS}\xspace}

\newcommand{\iso}[2]{\protect{\ensuremath{{}^{#1}\textrm{#2}}}\xspace}

\begin{document}


\title{Coherent elastic neutrino-nucleus scattering: Terrestrial and astrophysical applications}

%
%
\author[tamu]{M. Abdullah}
\author[TUW]{H.~Abele}
\author[mephi]{D. Akimov}
\author[MPP]{G.~Angloher}
\author[fed,bel]{D. Aristizabal Sierra}
\author[lyon]{C. Augier}
\author[wisc]{A.B. Balantekin}
\author[queensUCanada]{L. Balogh}
\author[duke,tunl]{P. S. Barbeau} 
\author[uzh]{L. Baudis}
% ...

%
%

\address[tamu]{Mitchell Institute for Fundamental Physics and Astronomy and Department of Physics and Astronomy, Texas A\&M University, College Station, TX 77843, USA}
\address[TUW]{Atominstitut, Technische Universit\"at Wien, A-1020 Wien, Austria}
\address[mephi]{National Research Nuclear University MEPhI (Moscow Engineering Physics Institute), Moscow, 115409, Russian Federation}
\address[MPP]{Max-Planck-Institut f\"ur Physik, D-80805 M\"unchen, Germany}
\address[fed]{Universidad T\'ecnica
  Federico Santa Mar\'{i}a - Departamento de F\'{i}sica\\
  Casilla 110-V, Avda. Espa\~na 1680, Valpara\'{i}so, Chile}%
\address[bel]{IFPA, Dep. AGO, Universit\'e de Li\`ege, Bat B5, Sart
  Tilman B-4000 Li\`ege 1, Belgium}%
\address[lyon]{Univ Lyon, Université Lyon 1, CNRS/IN2P3, IP2I-Lyon, F-69622, Villeurbanne, France}
\address[wisc]{Department of Physics, University of Wisconsin, Madison, WI 53706 USA}
\address[queensUCanada]{Department of Physics, Engineering Physics \& Astronomy, Queen's University, Kingston, Ontario K7L 3N6, Canada}
\address[duke]{Department of Physics, Duke University, Durham, NC 27708, USA}
\address[tunl]{Triangle Universities Nuclear Laboratory, Durham, NC 27708, USA}
\address[uzh]{Physik-Institut, University of Zurich, Winterthurerstr. 190, 8057 Zurich, Switzerland}
% ...

\date{\today}

\begin{abstract}
Coherent elastic neutrino-nucleus scattering (CE$\nu$NS) is a process in which neutrinos scatter on a nucleus which acts as a single particle.  Though the total cross section is large by neutrino standards, CE$\nu$NS has long proven difficult to detect, since the deposited energy into the nucleus is $\sim$ keV. In 2017, the COHERENT collaboration announced the detection of CE$\nu$NS using a stopped-pion source with CsI detectors, followed up the detection of CE$\nu$NS using an Ar target. The detection of CE$\nu$NS has spawned a flurry of activities in high-energy physics, inspiring new constraints on beyond the Standard Model (BSM) physics, and new experimental methods. The CE$\nu$NS process has important implications for not only high-energy physics, but also astrophysics, nuclear physics, and beyond. This whitepaper discusses the scientific importance of CE$\nu$NS, highlighting how present experiments such as COHERENT are informing theory, and also how future experiments will provide a wealth of information across the aforementioned fields of physics. 
\end{abstract}

\maketitle

\tableofcontents

\section{Introduction}
\par Neutrinos have proven valuable in elucidating the structure of the Standard Model (SM) of particle physics. Though the SM provides the framework describing how neutrinos interact with leptons and quarks through weak interactions, the SM does not answer fundamental questions about neutrinos. 
For example, how are neutrino masses generated? Are neutrinos Dirac or Majorana particles? 
Neutrinos provide both direct evidence for physics beyond the SM and a pathway to search for new physics. 

\par Neutrinos have been detected from many terrestrial and astrophysical sources, and across a large range of energy scales. Nuclear reactors produce neutrinos with $\gtrsim$ MeV energies, and accelerator sources produce neutrinos with energies $\gtrsim$ GeV. Neutrinos from astrophysical sources have been detected over an even larger energy range, from $\sim$ MeV up to PeV. Dating back to the detection of neutrinos from the Sun~\cite{Davis:1968cp}, from SN 1987A~\cite{Hirata:1987hu,Bionta:1987qt}, and more recently from a more distant and energetic source~\cite{IceCube:2018cha}, each detection has seemingly opened up a new window into the Universe.  

\par The low-energy MeV-scale has been especially important in elucidating the neutrino properties, such as their mass differences and mixing properties. At these energies, there is a unique complementarity in studies of neutrinos from terrestrial and astrophysical sources. Neutrinos from the Sun provide a direct probe of the nuclear burning process in the interior of stars~\citep{Robertson:2012ib}, and have been used in combination with long-baseline reactors~\cite{Eguchi:2002dm} to establish the LMA-MSW solution to the Solar neutrino problem. Atmospheric neutrinos have been used to establish the vacuum-oscillation solution to the angular dependence of the neutrino flux ~\citep{Fukuda:1998mi}.

\par At MeV energies, neutrinos have been detected via several distinct interaction channels. These channels include neutrino-electron elastic scattering $(\nu + e \rightarrow \nu + e)$, as well as neutral and charged current inelastic interactions on nucleons and nuclei~\cite{Formaggio:2013kya}. In the former case, detectors identify the energy in the outgoing electron, whereas in the latter case detectors identify either an electron produced in a charged current interaction, or an MeV gamma in the case of inelastic nuclear interactions. In the particular case of inverse-beta interaction ($\bar \nu_e + p \rightarrow e^+ + n$), both the outgoing positron and neutron may be detected. In the MeV energy regime, neutrino interactions transition from being described by point-like interactions with fundamental particles and interactions with constituent particles with the nucleus. 



Coherent elastic neutrino-nucleus scattering (\cevns) is a process in which neutrinos scatter on a nucleus which acts as a single particle. Within the SM, \cevns is fundamentally described by the neutral current interaction of neutrinos and quarks, and due to the nature of SM couplings it is proportional to the neutron number squared~\cite{Freedman:1973yd}. Though the total cross section is large by neutrino standards, \cevns has long proven difficult to detect, since the deposited energy into the nucleus is $\sim$ keV. In 2017, the COHERENT collaboration announced the detection of \cevns using a stopped-pion source with a CsI{[Na]} scintillating crystal detector~\cite{Akimov:2017ade}. This was followed up by the detection of \cevns with a single-phase liquid argon target~\cite{COHERENT:2020iec}, and with a larger exposure of CsI{[Na]}~\cite{Akimov:2021dab}. 

The detection of \cevns has motivated a flurry of theoretical activity in high-energy physics, inspiring new constraints on beyond the Standard Model (BSM) physics. It has motivated the development of larger scale detectors and technology to extend current detector sensitivity into lower, sub-keV scale energy regimes. The \cevns process has important implications for not only high-energy physics, but also astrophysics, nuclear physics, and beyond.

\par In addition to providing a new channel for detection of neutrinos, there are many interesting physics applications of \cevns-based experiments. One such example is the search for low-mass, $\lesssim$ GeV-scale dark matter particles. Since traditional WIMP direct dark matter detection searches lose sensitivity for WIMPs around a few GeV, \cevns-like experiments provide an important method to probe low-mass dark matter. These experiments are complementary to on-going experiments that are searching for dark matter at this mass scale. Beyond dark matter, \cevns-like experiments may be deployed to study long-sought-after particles such as axions. 

\par The goal of this whitepaper is to highlight the broad theoretical and experimental implications of the \cevns process. Section~\ref{sec:SM} discusses the calculation of the \cevns cross section in the SM, including a discussion of nuclear effects. Section~\ref{sec:terrestrial} discusses the terrestrial sources that are now being deployed to detect \cevns, and Section~\ref{sec:astrophysical} discusses astrophysical MeV neutrino sources. Section~\ref{sec:bsm} discussed the application for \cevns to physics beyond the Standard Model, focusing on non-standard neutrino interactions (NSI) and sterile neutrinos. Section~\ref{sec:experiments} reviews the on-going and future experiments around the world that are being utilized to detect~\cevns. Finally, Section~\ref{sec:connections} outlines broader connections to the US neutrino and dark matter physics program.

\section{\cevns in the Standard Model}
\label{sec:SM} 

\cevns is a neutral-current process that arises when the momentum transfer in the neutrino-nucleus interaction is less than the inverse of the size of the nucleus. 
In the SM, the interaction  is mediated by the $Z$-boson, with its vector component leading to the coherent enhancement~\cite{Freedman:1973yd}. As reference point, we first write the cross section in the form
\begin{align}
\frac{\text{d}\sigma}{\text{d}T}=\frac{G_F^2M}{4\pi}\bigg(1-\frac{M T}{2E_\nu^2}\bigg)Q_\text{w}^2\big[F_\text{w}(q^2)\big]^2\,,
\label{eq:SMcrosssection} 
\end{align}
where $G_F$ is the Fermi constant, $T=E_R=q^2/(2M)=E_\nu-E_\nu'$ is the nuclear recoil energy (taking values in $[0,2E_\nu^2/(M+2E_\nu)]$), $F_\text{w}(q^2)$ is the weak form factor, $M$ is the mass of the target nucleus, and $E_\nu$ ($E_\nu'$) is the energy of the incoming (outgoing) neutrino. The tree-level weak charge is defined by 
\begin{align}
Q_\text{w}=Z \big(1-4\sin^2\theta_W\big)-N\,,
\end{align}
with proton number $Z$, neutron number $N$, and weak mixing angle $\sin^2 \theta_W$. To first approximation, the weak form factor $F_\text{w}(q^2)$ depends on the nuclear density distribution of protons and neutrons. In the coherence limit $q^2\to 0$ it is normalized to $F_\text{w}(0)=1$, with the coherent enhancement of the cross section reflected by the scaling with $N^2$ via the weak charge, given the accidental suppression of the proton weak charge $Q_\text{w}^p\ll1$ (see Eq.~\eqref{eq:weakcharges} below). Consequently, this implies that \cevns is mainly sensitive to the neutron distribution in the nucleus.     

In writing the cross section as in Eq.~\eqref{eq:SMcrosssection} a number of subtleties have been ignored: subleading kinematic effects, axial-vector contributions, form-factor effects besides the density distributions, and radiative corrections. In the subsequent sections, some of these effects are addressed in more detail, see also Refs.~\cite{Hoferichter:2020osn,neutrino_WP}.  

\subsection{Structure of the Standard-Model contribution}

The quark-level interactions in the SM are
\begin{equation}
\label{Lagr_SM}
{\mathcal{L}}^\text{SM}=-\sqrt{2}G_F\sum_{q=u,d,s}\Big(C_q^V\bar\nu\gamma^\mu P_L\nu \,\bar q\gamma_\mu q
+C_q^A\bar\nu\gamma^\mu P_L\nu \,\bar q\gamma_\mu\gamma_5 q\Big)\,,
\end{equation}
with $P_L=(1-\gamma_5)/2$ and tree-level Wilson coefficients
\begin{align}
\label{Wilson_SM}
C_u^V=\frac{1}{2}\bigg(1-\frac{8}{3}\sin^2\theta_W\bigg)\,,\qquad
C_d^V=C_s^V=-\frac{1}{2}\bigg(1-\frac{4}{3}\sin^2\theta_W\bigg),\qquad
C_u^A=-C_d^A=-C_s^A=-\frac{1}{2}\,.
\end{align} 
The vector operator gives rise to the coherent contribution quoted in Eq.~\eqref{eq:SMcrosssection}, while the axial-vector operator adds an additional contribution that is not coherently enhanced. Including the dominant kinematic corrections, the cross section can be written in the form
\begin{equation}
\label{CEvNS_SM}
\frac{\text{d} \sigma}{\text{d} T}=\frac{G_F^2 M}{4\pi}\bigg(1-\frac{M T}{2E_\nu^2}-\frac{T}{E_\nu}\bigg)Q_\text{w}^2\big[F_\text{w}(q^2)\big]^2
+\frac{G_F^2M}{4\pi}\bigg(1+\frac{M T}{2E_\nu^2}-\frac{T}{E_\nu}\bigg)F_A(q^2)\,,
\end{equation}
with an axial-vector form factor $F_A(q^2)$~\cite{Hoferichter:2020osn}. This contribution vanishes for nuclei with even number of protons and neutrons, which have spin-zero ground states. 

Moving from the quark-level interactions in Equation~\eqref{Lagr_SM} to the neutrino-nucleus cross section in Eq.~\eqref{CEvNS_SM} involves a two-step process~\cite{Hoferichter:2020osn}. In the first step, hadronic matrix elements are required to obtain the matching to single-nucleon operators, i.e., vector and axial-vector form factors of the nucleon, respectively. For the vector operators, the normalization is determined via the valence-quark content, leading to the relations
\begin{equation}
\label{eq:weakcharges}
Q_\text{w}^p=2(2C_u^V+C_d^V)=1-4\sin^2\theta_W\,,\qquad 
Q_\text{w}^n=2(C_u^V+2C_d^V)=-1\,,
\end{equation}
while the $q$-dependent corrections, expressed in terms of radii and magnetic moments, are subsumed into the weak form factor $F_\text{w}(q^2)$. Similarly, $F_A(q^2)$ depends on the axial charges and radii of the nucleon. 
In the second step, the nuclear responses need to be derived from a multipole expansion~\cite{Serot:1978vj,Donnelly:1978tz,Donnelly:1979ezn,Serot:1979yk,Walecka:1995mi}, in which the leading contribution can be interpreted in terms of the proton and neutron density distributions. 
The relations~\eqref{eq:weakcharges} hold at tree level in the SM, with radiative corrections discussed in Sec.~\ref{sec:radiative}.

Instead of writing the \cevns cross section in terms of the recoil energy, as in Eq.~\eqref{eq:SMcrosssection}, the cross section may also be expressed in terms of the direction of the recoil, converting the recoil to an angular spectrum. In practice, a detector may provide a measurement of both recoil energy and direction at once, in which case the scattering rate would be expressed as a function of both variables, $\text{d}^2R/(\text{d}\Omega_R \text{d}E_R)$, where the angles are those of the scattered nucleus measured with respect to the incident neutrino direction. This quantity has been referred to in the literature as the Momentum Spectrum~\cite{Gondolo:2002np} and as the Directional Recoil Spectrum (DRS)~\cite{Abdullah:2020iiv}. It can be written as
\begin{equation}
\frac{\text{d}^2R}{\text{d}\Omega_R\text{d}E_R}=\frac{1}{2\pi}
\left .\frac{\text{d}\sigma}{\text{d}E_R}\right|_{E_\nu=\varepsilon}\,
\frac{\varepsilon^2}{E_\nu^\text{min}}
\left .\frac{\text{d}\Phi}{\text{d}E_\nu}\right|_{E_\nu=\varepsilon}\,,
\end{equation}
where $\text{d}\Phi/\text{d}E_\nu$ is the differential neutrino flux, $E_\nu^\text{min} = \sqrt{M E_R/2}$, and 
\begin{equation}
\frac{1}{\varepsilon}=\frac{\cos \theta_R}
{E_\nu^\text{min}}-\frac{1}{M}\,.
\end{equation}
To switch variables directly between $E_R$ and $\Omega_R$ one can use the following relation and the associated Jacobian:
\begin{equation}
E_R=\frac{2ME_\nu^2\cos^2\theta_R}{(E_\nu+M)^2-E_\nu^2\cos^2\theta_R}\,.
\end{equation}
The directional and energy double differential cross section can be written by noting that the scattering has azimuthal symmetry about the incoming neutrino direction. Integrating over outgoing nuclear recoil energy gives
\begin{equation}
\label{dsigmadOmega}
\frac{\text{d}\sigma}{\text{d}\Omega_R} = \frac{G_F^2}{16\pi^2}
Q_\text{w}^2
E_\nu (1+\cos \theta_R)
\big[F_\text{w}(q^2)\big]^2 \,, 
\end{equation}
where the angle is defined as $\text{d}\Omega_R = 2\pi \cos \theta_R \text{d}\theta_R$, and $\theta_R$ is the scattering angle between the direction of the incoming and outgoing neutrino. 

\subsection{Nuclear and hadronic physics} 
\label{sec:Nuclear}

Due to the suppression of the weak charge of the proton, the most important nuclear response required for the interpretation of \cevns experiments is related to the neutron distribution. While the charge density of nuclei has been probed extensively in elastic electron scattering experiments~\cite{Hofstadter:1956qs,DeJager:1987qc, Fricke:1995zz, Angeli:2013epw}, the neutron density distributions are hard to determine. Precise experimental data exist for observables that are sensitive to the neutron density distribution or the neutron skin, such as the nuclear dipole polarizability~\cite{Tamii:2011pv,Rossi:2013xha,Hashimoto:2015ema,Birkhan:2016qkr}, but efforts using hadronic probes require a careful analysis of model-dependent uncertainties (see, e.g., Ref.~\cite{Thiel:2019tkm}). In contrast, electroweak processes such as parity-violating electron scattering (PVES)~\cite{Donnelly:1989qs} and \cevns
have long been considered as clean probes of the neutron densities. Both of which, though long considered experimentally challenging, have become a reality in recent years~\cite{Abrahamyan:2012gp, Horowitz:2012tj,Kumar:2020ejz,PREX:2021umo}. 


The observation of \cevns can therefore further provide important nuclear structure information through the determination of the weak form factor, which constrains the neutron density distribution and thus the neutron radius and the neutron skin, at least at low momentum transfers where the process remains coherent~\cite{Horowitz:2003cz,Patton:2012jr,Cadeddu:2017etk,Ciuffoli:2018qem,Payne:2019wvy,Yang:2019pbx,AristizabalSierra:2019zmy,Papoulias:2019lfi,Hoferichter:2020osn,Co:2020gwl,Coloma:2020nhf,VanDessel:2020epd}. 
These measurements complement PVES experiments not only due to 
additional data, but also due to different energy ranges and nuclear targets, which could be used to calibrate nuclear-structure calculations. Furthermore, improved measurements of the neutron skin would have important consequences for the equation of state of neutron-rich matter, which plays an essential role in understanding the structure and evolution of neutron stars~\cite{RocaMaza:2011pm,Tsang:2012se,Lattimer:2012xj,Hebeler:2013nza,Hagen:2015yea}.

However, arguably the most intricate aspect of nuclear-structure input concerns searches for physics beyond the SM (BSM). Without independent experimental information for the neutron responses, which, potentially apart from PVES, is difficult to obtain, \cevns cross sections provide constraints on the combination of nuclear responses and BSM effects. In fact, in order to derive BSM constraints beyond the level at which current nuclear-structure calculations constrain the neutron distribution, a combined analysis of multiple targets and momentum transfers is required to distinguish between nuclear structure and potential BSM contributions. To do so, a detailed understanding of the nuclear responses is prerequisite.


\begin{figure}[t]
	\begin{center}
		\includegraphics[width=0.48\textwidth,clip]{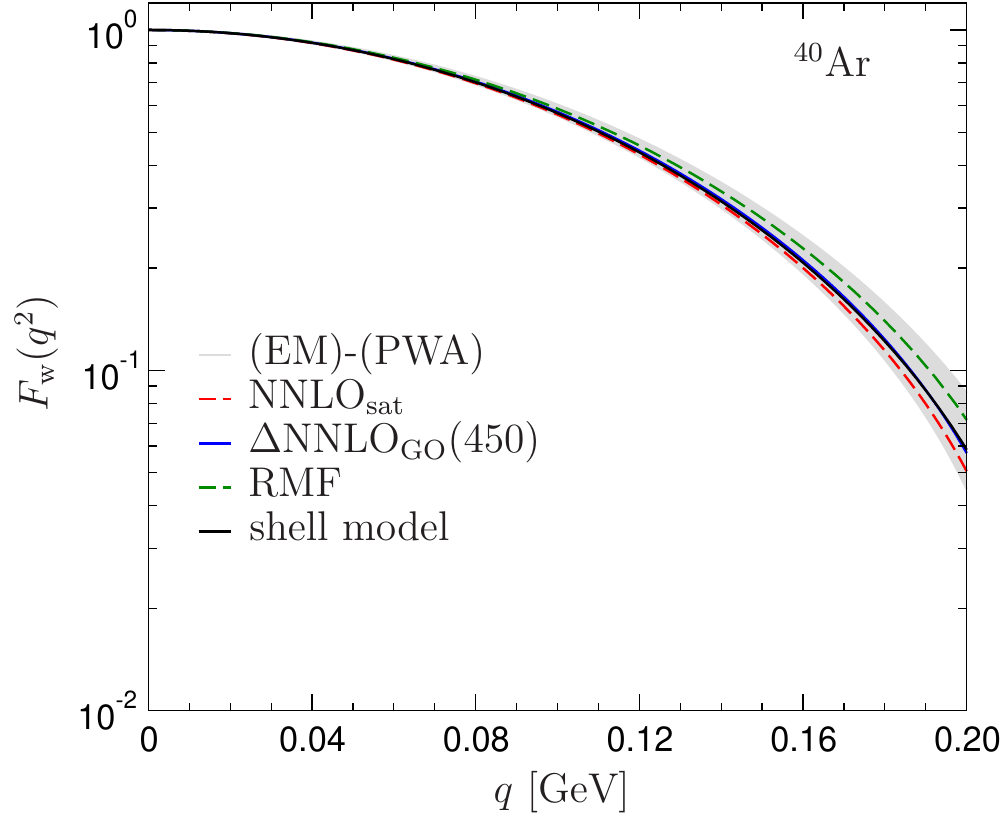}
	\end{center}
	\caption{Theoretical predictions for the weak form factor of $^{40}$Ar, from relativistic mean-field methods~\cite{Yang:2019pbx}, coupled-cluster~\cite{Payne:2019wvy}, and shell-model~\cite{Hoferichter:2020osn} calculations. The curves/bands labeled (EM)-(PWA), NNLO$_\text{sat}$, and $\Delta$NNLO$_\text{GO}$(450) refer to the chiral interactions considered in Ref.~\cite{Payne:2019wvy}. Figure adapted from Ref.~\cite{Hoferichter:2020osn}.}
	\label{fig:Fweak_Ar}
\end{figure}


Traditionally, the weak form factor 
\begin{equation}
\label{Eq:Fw}
F_\text{w}(q^2) = \frac{1}{Q_\text{w}} \left[ Z Q_\text{w}^p F_{p}(q^2) +N Q_\text{w}^n F_n(q^2)\right]
\end{equation}
has been modeled in terms of proton and neutron densities 
\begin{equation}
\label{Eq:Fn}
F_n(q^2) = \frac{4\pi}{N} \int \text{d}r~r^2~\frac{\sin(qr)}{qr}~\rho_n(r)\,,\qquad 
F_p(q^2) = \frac{4\pi}{Z} \int \text{d}r~r^2~\frac{\sin(qr)}{qr}~\rho_p(r)\,,
\end{equation}
where $\rho_n(r)$ and $\rho_p(r)$ are neutron and proton density distributions normalized to the neutron and proton numbers. 
Phenomenological form factors, such as Helm~\cite{Helm:1956zz} and Klein-Nystrand~\cite{Klein:1999qj}, are 
based on empirical fits to elastic electron scattering data, and similar parameterizations are assumed for the neutron form factor.
In the Helm approach~\cite{Helm:1956zz}, the nucleon distribution is given by the convolution of a uniform density with radius $R_0$ and a Gaussian profile with width $s$, the surface thickness. The resulting form factor is
\begin{equation}
F_{\text{Helm}}(q^2) = \frac{3 j_1(qR_0)}{qR_0} e^{-q^2s^2/2}\,,
\end{equation}
where $j_1(x)$ is the spherical Bessel function of order one. The Klein-Nystrand approach~\cite{Klein:1999qj} relies on a surface-diffuse distribution that results from folding a short-range Yukawa potential with range $a_k$ over a hard sphere distribution with radius $R_A$. The resulting form factor becomes
\begin{equation}
F_{\text{KN}}(q^2) = \frac{3 j_1(qR_A)}{qR_A} \left[\frac{1}{1+q^2a_k^2} \right]\,.
\end{equation}
In both cases, it should be stressed that these parameterizations need to assume a value for the neutron radius---related to $R_0$ or $R_A$---and only try to capture the leading nuclear responses, with the neutron distribution largely unconstrained. Actual nuclear-structure calculations of the nuclear responses are based on relativistic mean-field methods~\cite{Horowitz:2003cz,Yang:2019pbx},
nonrelativistic energy-density functionals~\cite{Patton:2012jr,Co:2020gwl,VanDessel:2020epd}, shell-model calculations~\cite{Hoferichter:2016nvd,Hoferichter:2018acd,Hoferichter:2020osn}, and, for argon, 
a first-principles calculation using coupled-cluster theory~\cite{Payne:2019wvy}.

Retaining all responses that at least display some degree of coherent enhancement, the weak form factor receives further contributions, e.g., related to finite-size effects and spin-orbit interactions. 
Further corrections could be expected from two-body currents, but for the relevant responses such contributions only start at loop level in the chiral expansion~\cite{Hoferichter:2020osn}. Figure~\ref{fig:Fweak_Ar} compares several predictions for argon's $F_\text{w}(q^2)$, whose spread indicates the accuracy with which nuclear responses can currently be calculated.


\subsection{Radiative corrections}
\label{sec:radiative}

The relation~\eqref{eq:weakcharges} for the weak charges holds true at tree-level, in which case $Q_\text{w}^{p,n}$ are flavor universal and apply both to neutrino and electron scattering. Once including radiative corrections, process- and flavor-dependent contributions arise, in such a way that separate weak charges need to be defined. 
For \cevns, the corresponding radiative corrective have been studied in Refs.~\cite{Barranco:2005yy,Erler:2013xha,Tomalak:2020zfh}. Keeping the decomposition $Q_\text{w}=Z Q_\text{w}^p+N Q_\text{w}^n$, one has from Ref.~\cite{Erler:2013xha}
\begin{align}
 Q_\text{w}^{\nu_e, p}&=0.0766\,, & Q_\text{w}^{\nu_\mu, p}&=0.0601\,, &
 Q_\text{w}^{\nu_\tau, p}&=0.0513\,,\notag\\
 Q_\text{w}^{\nu_\ell, n}&=-1.0233\,,
\end{align}
i.e., only $Q_\text{w}^{\nu_\ell, p}$ becomes flavor dependent. These values are in agreement with Ref.~\cite{Tomalak:2020zfh}
\begin{align}
    Q_\text{w}^{\nu_e, p}&=0.0747(34)\,, & Q_\text{w}^{\nu_e, p}-Q_\text{w}^{\nu_\mu, p}&=0.01654\,, & Q_\text{w}^{\nu_\mu, p}-Q_\text{w}^{\nu_\tau, p}&=0.00876\,,\notag\\
    Q_\text{w}^{\nu_\ell, n}&=-1.02352(25)\,. 
\end{align}
The main difference between Refs.~\cite{Erler:2013xha,Tomalak:2020zfh} concerns the treatment of the light-quark loops in $\gamma$--$Z$ mixing diagrams, which lead to non-perturbative effects that have been absorbed into $Q_\text{w}^{\nu_\ell, p}$. 

The consequences of process-dependent corrections become apparent when comparing to the SM values for the weak charges probed in electron scattering~\cite{Erler:2013xha,Zyla:2020zbs}
\begin{equation}
  Q_\text{w}^{e, p}=0.0710\,,\qquad   Q_\text{w}^{e, n}=-0.9891\,,
\end{equation}
which include further corrections ($\gamma Z$ box diagrams and axial-current renormalization) that do not play a role in \cevns.

\section{Terrestrial sources}
\label{sec:terrestrial} 

\par In this section we discuss terrestrial neutrino sources that are deployed to detect CE$\nu$NS. In Section~\ref{sec:experiments} below we provide more detailed information on current and forthcoming experimental efforts that use these sources. 

\subsection{Stopped-pion beams} 
\par Spallation sources produce $\pi^+$ and $\pi^-$ though proton collisions with nuclei. Most of the $\pi^-$ that are produced are captured by nuclei, and therefore do not decay to produce neutrinos. On the other hand, the $\pi^+$ lose energy and decay at rest, $\pi^+ \rightarrow \mu^+ \nu_\mu$, to produce mono-energetic muon neutrinos with energy 30 MeV. From the subsequent decay of $\mu^+$ at rest, $\bar{\nu}_\mu$ and $\nu_e$ are produced with a Michel energy spectrum. Due to the decay lifetime, $\bar{\nu}_\mu$ and $\nu_e$ from muon decays are delayed relative to the 30 MeV $\nu_\mu$ produced from the prompt pion decay. The spectral functions are given by
\begin{eqnarray}
  \label{eq:spectra}
  \mathcal{F}_{\nu_\mu}(E_\nu)&=&\frac{2m_\pi}{m_\pi^2-m_\mu^2}
                                \delta\left(1-\frac{2E_\nu m_\pi}{m_\pi^2-m_\mu^2}\right)\ ,
                                \nonumber\\
  \mathcal{F}_{\nu_e}(E_\nu)&=&\frac{192}{m_\mu}\left(\frac{E_\nu}{m_\mu}\right)^2
                              \left(\frac{1}{2}-\frac{E_\nu}{m_\mu}\right)\ ,
                              \nonumber\\
  \mathcal{F}_{\bar\nu_\mu}(E_\nu)&=&\frac{64}{m_\mu}\left(\frac{E_\nu}{m_\mu}\right)^2
                                    \left(\frac{3}{4}-\frac{E_\nu}{m_\mu}\right)\ 
                                    \nonumber.
\end{eqnarray}
For a pion-at-rest source $E_\nu^\text{max}=m_\mu/2$ where $m_\mu = 105.65$ MeV is the muon mass. 

\begin{figure}
    \centering
    \includegraphics[width=0.48\textwidth]{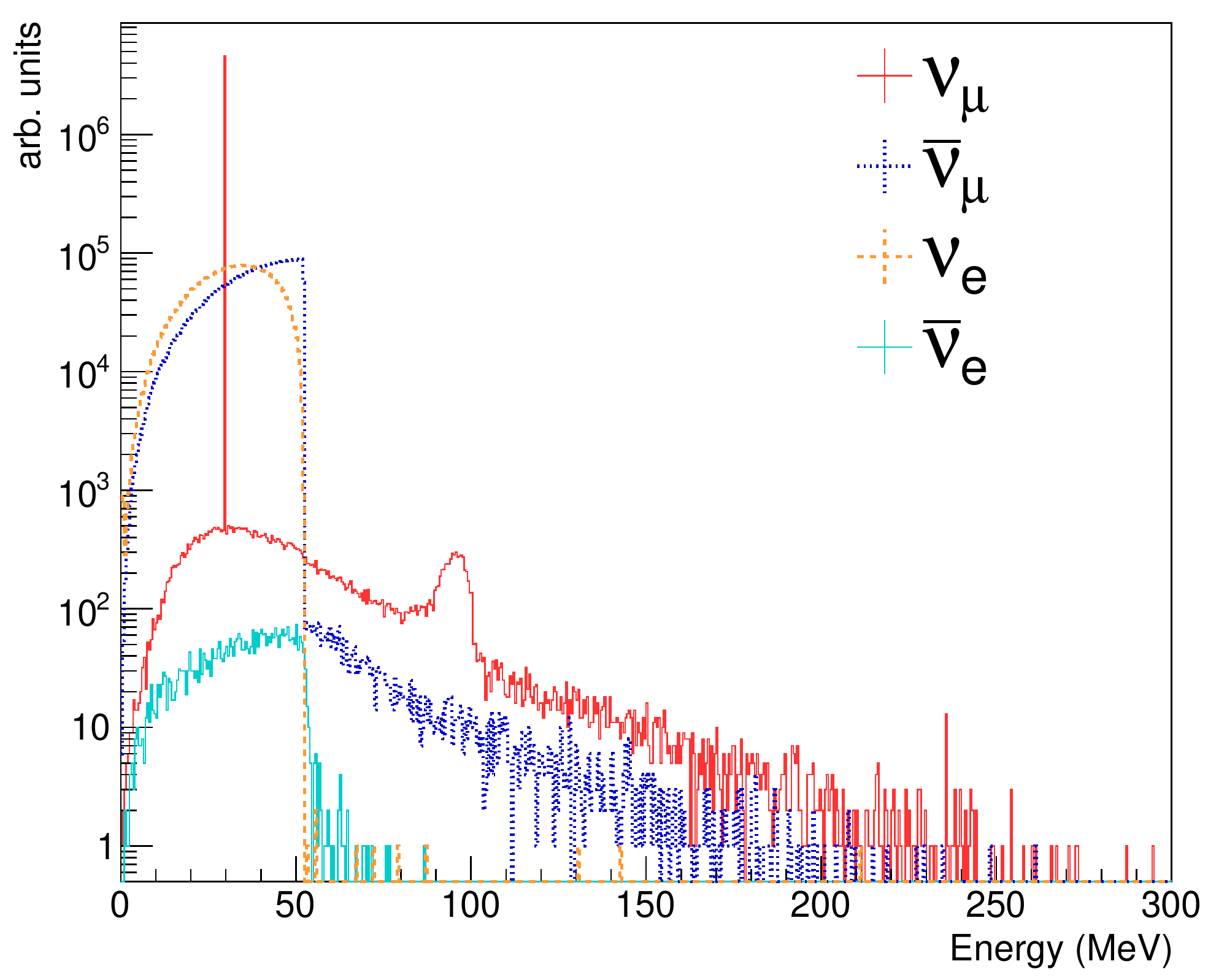}
    \includegraphics[width=0.48\textwidth]{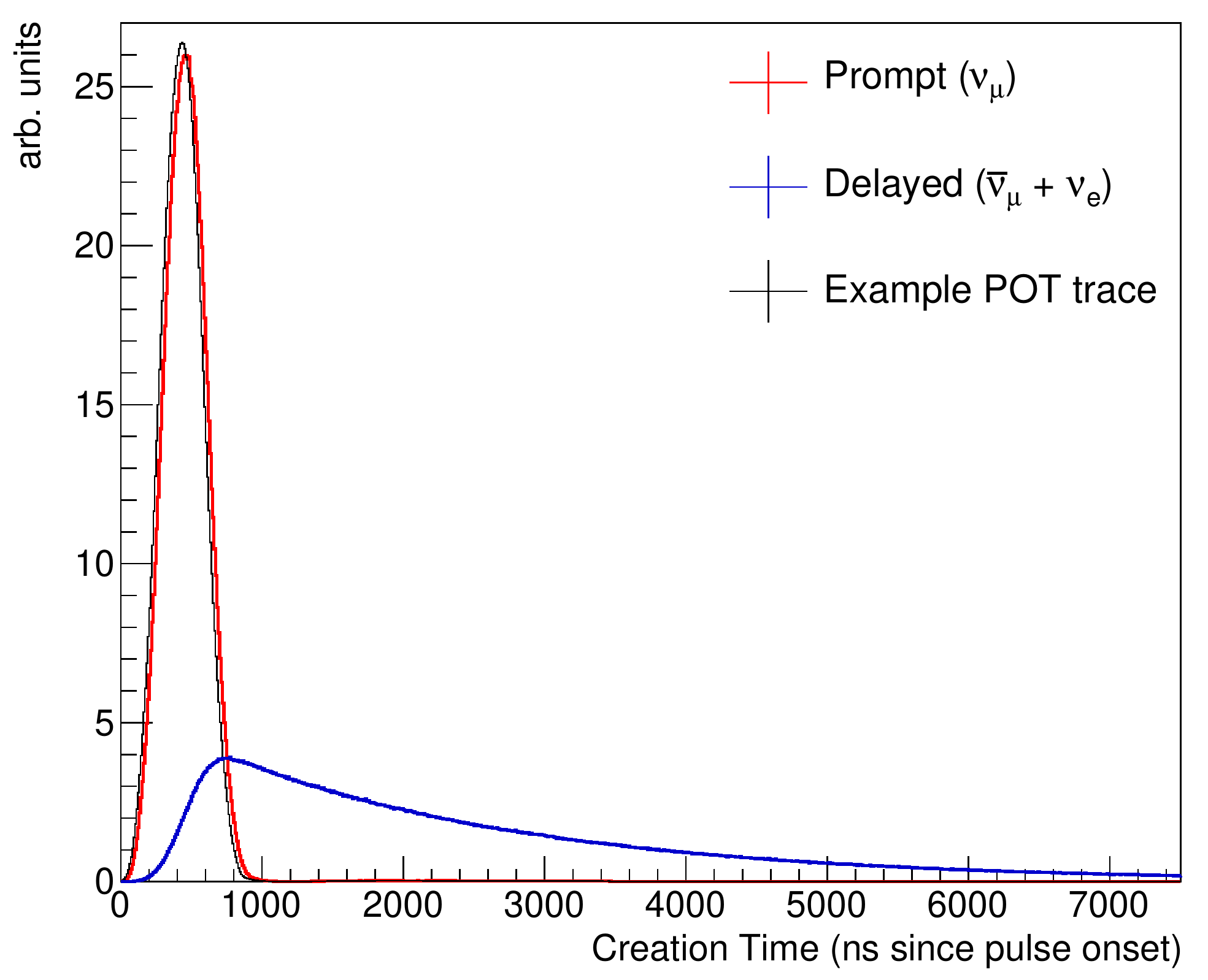}
    \caption{Distributions of neutrino energy (left) and creation time (right) from the SNS. Pion decay at rest components are prominent at low energies, with the high energy tails resulting from pion decay in flight.  The SNS beam spill (350 ns FWHM) allows for flavor separation due to the delayed decay of the muon.  Figures reproduced from~\cite{snsFlux2021}.
    \label{fig:sns}}
\end{figure}

\subsection{Reactors}
\par
\begin{figure}
    \centering
    \includegraphics[width=0.50\textwidth]{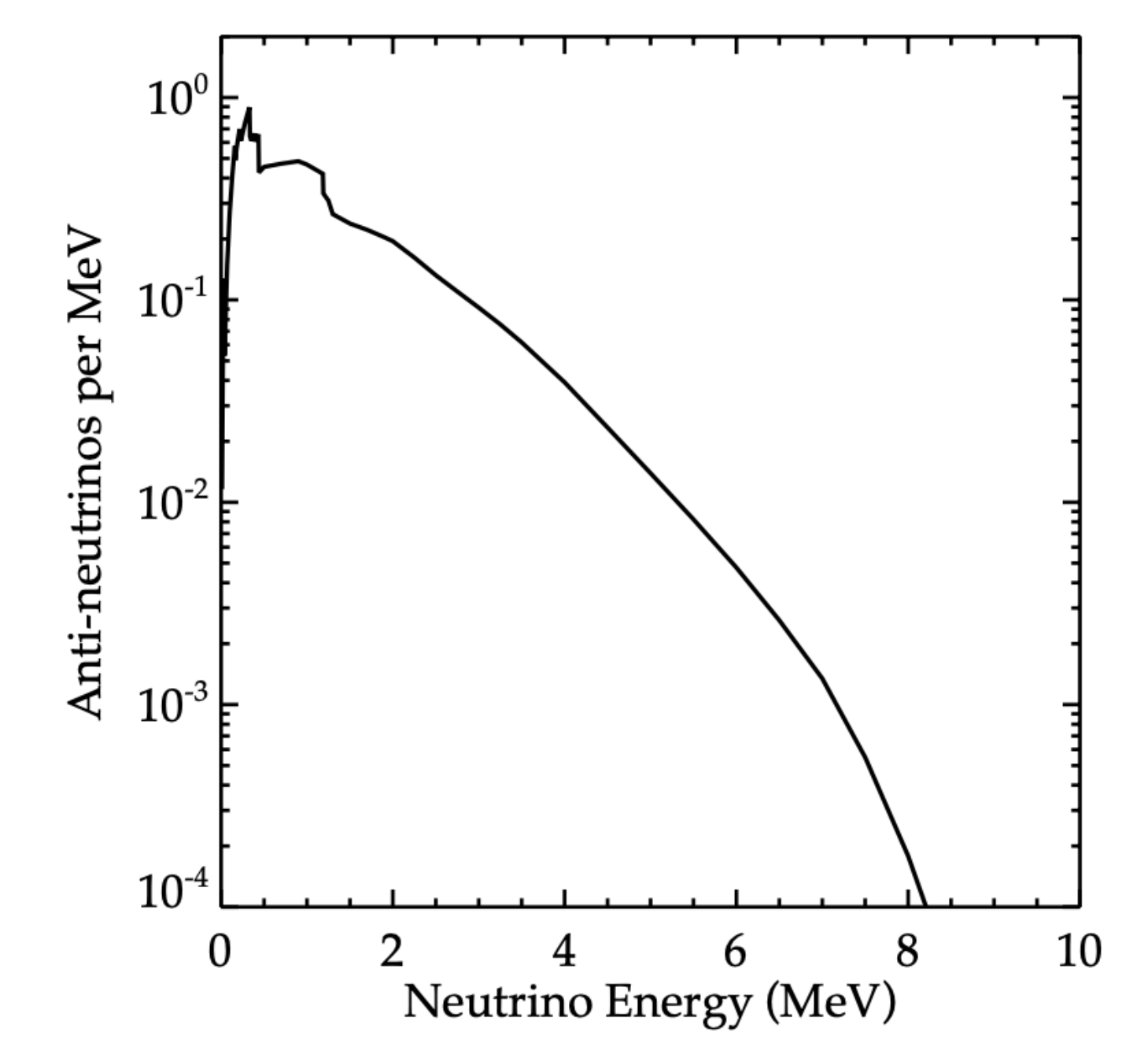}
    \caption{Electron anti-neutrino spectrum from nuclear reactors.
    \label{fig:reactor}}
\end{figure}

 Nuclear reactors have long been purposed as copious sources of electron anti-neutrinos. Neutrinos from reactors have been detected using the inverse beta decay reaction, $\bar{\nu}_e + p \rightarrow e^+ + n$, by observing both the outgoing positron and coincident neutron. There are four isotopes whose fission produce neutrinos above the inverse beta decay threshold: $^{235}U$, $^{241}P$, $^{239}P$, and $^{238}U$. The neutrino flux is determined from the power produced by the reactor, with theoretical uncertainties on the reactors fluxes estimated in Refs.~\cite{Huber:2011wv,Mueller:2011nm}. The calculated electron anti-neutrino spectrum from reactors is shown in Figure~\ref{fig:reactor}. 

\par The characteristic neutrino energy is $\lesssim 1$ MeV, which is nearly an order of magnitude less than the neutrinos produced by accelerator sources. Due to these low energies, the coherence condition for the recoil is largely preserved over the entire reactor energy regime, so that there is no dependence on the internal structure of the nucleus~\cite{Bednyakov:2018mjd}. 

\subsection{$^{51}$Cr}
$^{51}$Cr is an electron-capture decaying isotope with a half-life of 27.7 days. The neutrino spectrum consists of four monochromatic lines, the most energetic of which are at 747 keV (81\%) and 752 keV (9\%). These lines may be exploited for CE$\nu$NS~\cite{Bellenghi:2019vtc}.  

\subsection{Geo-neutrinos}

Geo-neutrinos emitted from radioactive decays of $^{238}$U, $^{232}$Th and $^{40}$K within Earth can provide unique insight into its interior, formation and central engine. While the geo-neutrino flux is suppressed, CE$\nu$NS has the potential to explore beyond the 1.8~MeV kinematic threshold limiting the conventional IBD signal and thus detect the $^{40}$K geoneutrinos~\cite{Gelmini:2018gqa}.

\subsection{Next-generation neutrino beams}
The Long-Baseline Neutrino Facility (LBNF) beamline at Fermilab may also be used to study CE$\nu$NS~\citep{AristizabalSierra:2021uob}. The LBNF beam neutrinos are produced at a characteristic energy scale different than neutrinos from reactor or SNS sources. This provides an important new, third energy scale at which the CE$\nu$NS cross section can be studied.

Although energetic, the LBNF beamline can induce CE$\nu$NS and that the process can be measured,
provided the detector is sensitive to low recoil energies. The low-energy tail of the neutrino spectrum (on-axis) extends down to energies of order 50 MeV. 

\section{Astrophysical sources} 
\label{sec:astrophysical} 
\par In this section we discuss neutrinos from astrophysical sources, including the Sun, supernovae, and the atmosphere, that may be used for CE$\nu$NS detection. 

\subsection{Solar neutrinos} 
\par The field of Solar neutrinos is over a half-century old. The primary goal of these solar neutrino experiments is to measure the different components of the flux, and use these measurements to understand the physics of the solar interior. The first experiments utilized neutrino capture reactions on Cl, Ga to specifically study the electron neutrino component of the neutrino flux~\cite{Cleveland:1998nv,Abdurashitov:2002nt}. ``Real-time" solar neutrino kilo-ton scale water cherenkov experiments~\cite{Abe:2010hy,Aharmim:2011vm,Bellini:2011rx} measured neutrino-electron elastic scattering, with sensitivity to both electron and muon neutrino flavors. Combining all experimental data, the flux of the 8B component is $5.25 \times 10^6$ cm$^{-2}$ s$^{-1}$~\citep{Aharmim:2011vm}.

\par Borexino has measured the low-energy components of the solar neutrino flux, originating from the reactions $p + e^- + p$ (pep), ${}^7$Be, $p + p$ (pp), and CNO cycles~\cite{Bellini:2011rx,Bellini:2013lnn,Borexino:2017rsf,Bellini:2014uqa,Agostini:2020mfq}. The combination of all solar neutrino data with terrestrial experiments favor the LMA-MSW solution to neutrino flavor transformation from the Sun to the Earth. At low energies, $\lesssim 5$ MeV, vacuum oscillations describe the neutrino flavor transformation, and the electron neutrino survival probability is $\gtrsim 50\%$. At energies $\gtrsim 5$ MeV, matter-induced transformations describe the flavor transformation, with a corresponding survival probability of $\gtrsim 1/3$~\cite{Robertson:2012ib,Antonelli:2012qu}. 

\par Even with the tremendous theoretical and experimental progress in the field of solar neutrinos, there are still some outstanding questions that surround some of the data. For example, three experiments (Super-Kamiokande, SNO, and Borexino) that are sensitive to electron recoils from neutrino-electron elastic scattering find that at electron recoil energies of a few MeV, the data are $\sim 2\sigma$ discrepant relative to the prediction of the best-fitting LMA-MSA solution. This may be indicative of new physics~\citep{deHolanda:2010am}. In addition, the recent measurement of the solar mass-squared difference from solar neutrino data, in particular from the day-night Super-Kamiokande data~\cite{Abe:2016nxk}, is discrepant at the $\sim 2 \sigma$ level relative to that measured by KamLAND~\cite{Gando:2010aa}. This may be explained by novel physics in the neutrino sector~\cite{Liao:2017awz}.

\par Another outstanding question relates to how the measured flux informs the physics of the solar interior. Modeling of solar absorption spectra and heliosiesmology data suggests a lower abundance of metals in the solar core, i.e. a low-Z SSM~\cite{Asplund:2009fu}. This is in comparison to the previously-established high-Z SSM~\cite{Grevesse:1998bj}. Though some sets of solar neutrino data favor a high-Z SSM~\cite{Borexino:2017rsf}, a global analysis of all solar neutrino fluxes remains inconclusive~\cite{Bergstrom:2016cbh}.

\par
\begin{figure}[!htb]
\centering
\includegraphics[width=2.0in]{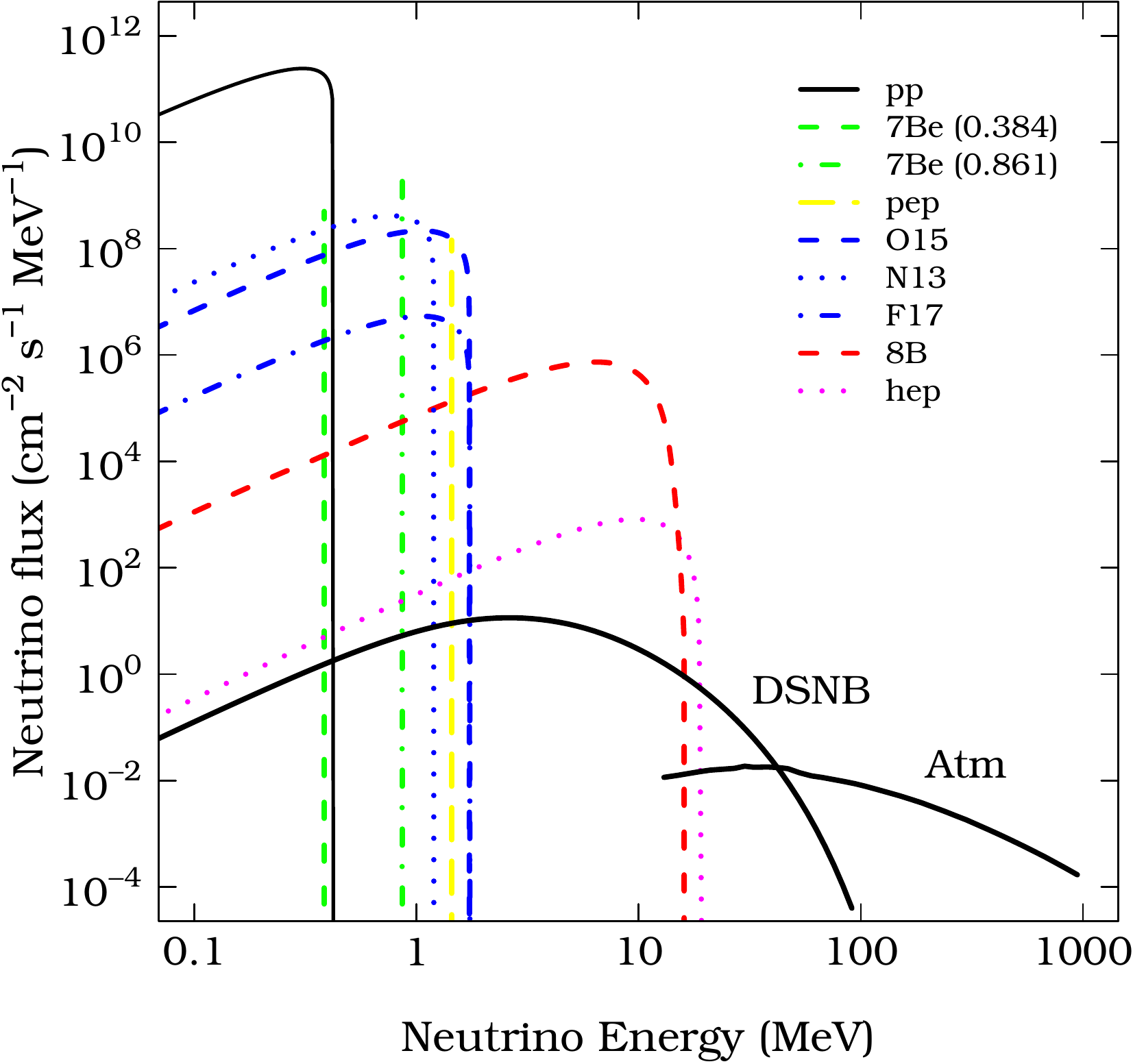}
\includegraphics[width=2.0in]{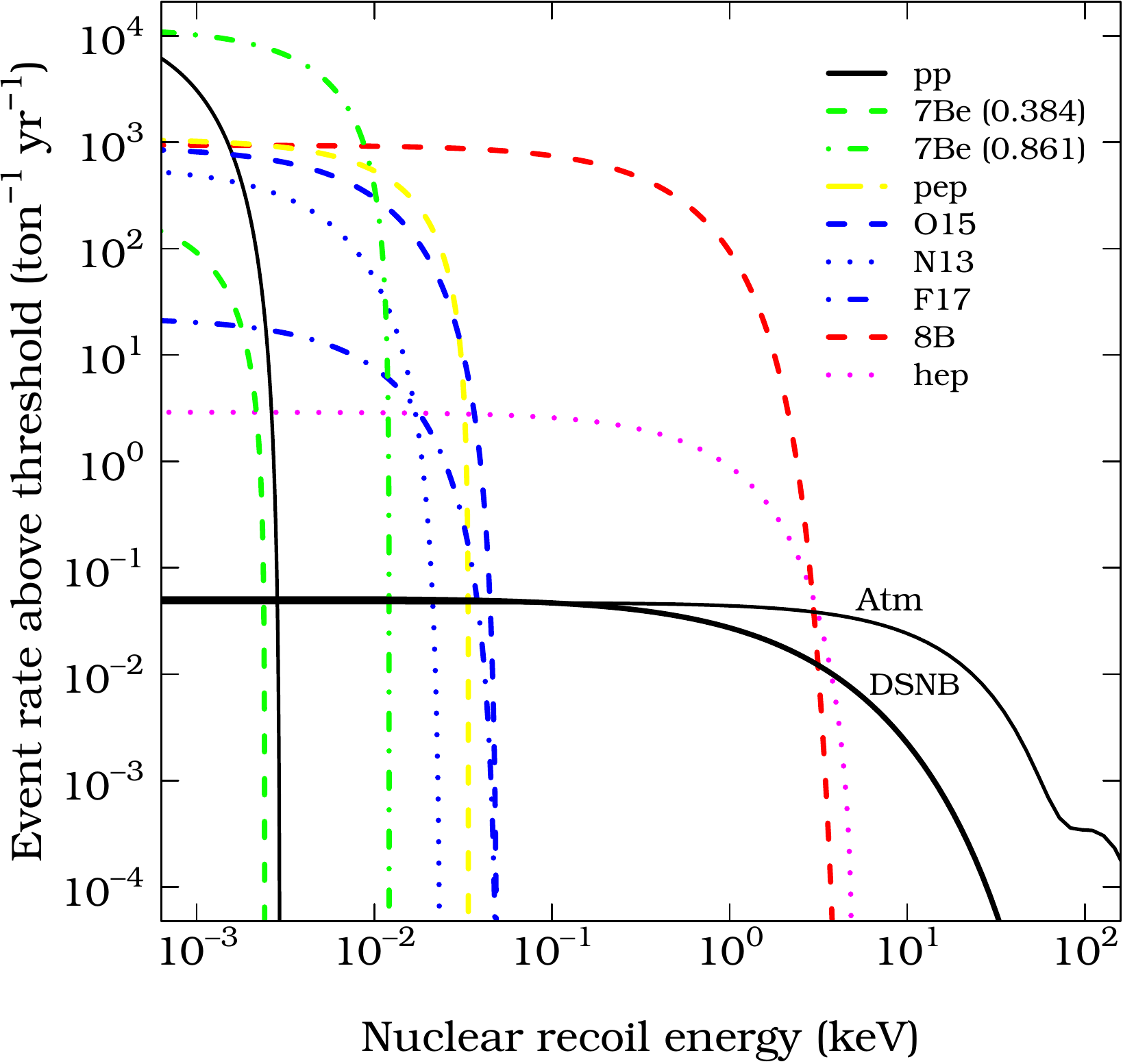}
\caption{Solar, atmospheric, and diffuse supernova neutrino fluxes. For the solar fluxes, each individual component is shown. For the atmospheric flux, shown is the sum of all flavor components. For the diffuse supernova neutrino fluxes, shown is the sum of all flavor components.} 
\label{fig:nuflux}
\end{figure}

 Figure~\ref{fig:nuflux} shows the solar neutrino fluxes, with the normalizations of each flux corresponding to those from the high-Z SSM. The pp and pep are relatively insensitive to the assumed solar metallicity model, while the $^{8}$B, $^{7}$Be, and the CNO components are much more sensitive to the solar metallicity model. 
There is a particularly large theoretical uncertainty ($\sim 15\%$) on the CNO neutrino flux, which has just recently been detected by Borexino. 

\subsection{Supernova neutrinos} 
\par The detection twenty neutrinos from 1987A confirmed that core-collapse supernova explosions carry away 99\% of the energy associated with the burst in neutrinos of all flavors. Neutrinos are expected to emerge from the core of a supernova with a nearly Fermi-Dirac spectrum, with temperatures for $\nu_e$, $\bar{\nu}_e$, and muon/tau flavors of $\sim 3,5,8$ MeV, respectively~\cite{Raffelt:2003en}. Though neutrino detectors have been operating in the three decades since SN 1987A, there has yet to be a detection from neutrinos from a Galactic supernova~\cite{Ikeda:2007sa}. The next supernova event in the Milky Way or in nearby galaxies is expected to provide unprecedented information on the physics of neutrino propagation from the SN core~\cite{Janka:2012wk}. For example, large water cherenkov detectors such as Super-Kamiokande will measure thousands of events, mostly through the charged-current inverse beta decay channel, and hundreds of events through various other elastic and inelastic channels~\cite{Scholberg:2012id}. 

\par A series of detectors are currently waiting for neutrinos from the next nearby supernova (see SNEWS section below). Detectors sensitive to CE$\nu$NS will play an important role in extracting information from supernova neutrinos. Through the CE$\nu$NS channel, and neutrinos flavors will be accessible~\citep{Horowitz:2003cz,Lang:2016zhv,Pattavina:2020cqc}. For many years, this has been recognized to provide important information on the nature of stellar collapse~\citep{Freedman:1977xn}. By measuring the mean neutrino energies, tens of events are likely enough to constrain the explosion energy of the supernova, and to reconstruct the supernova lightcurve. This will be possible with currently-operating ton-scale detectors, and with next generation, multi-ton scale detectors the extracted physics will rival that of more traditional neutrino detectors~\cite{Lang:2016zhv}. In addition to the well-established technology of Xe-based time projection chambers, Pb-based cryogenic detectors can offer a great advantage thanks to their high CE$\nu$NS cross-section, and their optimized small experimental volume. The newly proposed RES-NOVA experiment is aiming at achieving a precision, in the reconstruction of the main supernova parameters, comparable to the one of the currently running experiments, while operating a modular \textit{cm}-scale archaeological Pb-based cryogenic detector~\cite{RES-NOVA:2021gqp}. More-so, CE$\nu$NS can be effectively utilized to detect the pre-SN neutrino flux emitted prior to the SN onset, allowing the detectors to act as SN alarms and providing information about stellar evolution complimentary to IBD signal employed in conventional neutrino searches~\cite{Raj:2019wpy}.

\par In addition to the yield from a Galactic supernova event, future dark matter detectors may have sensitivity to the diffuse supernova neutrino background (DSNB)~\cite{Lunardini:2010ab,Beacom:2010kk}. The predicted DSNB flux is $\sim 6$ cm$^{-2}$ s$^{-1}$~\cite{Horiuchi:2008jz}, including contributions from all neutrino flavors. Though the DSNB has not been directly detected, there are strong upper bounds on the $\bar{\nu}_e$ component of the flux from Super-Kamiokande~\cite{Bays:2011si}. The best predictions for the flux of all flavors implies that dark matter detectors with exposures $\sim 100$ ton-year should be sensitive to the DSNB~\cite{Strigari:2009bq,Suliga:2021hek}. Analogously to supernova events and DSNB, CE$\nu$NS can probe explosions of supermassive stars and associated diffuse neutrino background~\cite{Munoz:2021sad}, which could be be related to the origin of supermassive black holes.

\subsection{Atmospheric neutrinos}
\par The collisions of cosmic rays in the atmosphere produce mesons and leptons across a wide range of energies. These heavy mesons and leptons decay to produce muon and electron neutrinos and antineutrinos. A precise determination of this atmospheric neutrino flux at the surface of the Earth depends on several factors, including the cosmic-ray flux at the top of the Earth's atmosphere, the propagation of the cosmic rays through the atmosphere, and the decay of the mesons and muons as they propagate though the atmosphere. Since the flavors of neutrinos that are produced in the decays are known, theoretical models accurately predict the ratio of the flavor components of neutrinos across all energies. However, the normalizations of the fluxes differ depending upon the theoretical input.  

\par Dating back several decades since their initial detection~\cite{,Achar:1965ova,Reines:1965qk}, many experiments have estimated the flux of atmospheric neutrinos over nearly the entire energy range which they are produced~\cite{Richard:2015aua,Aharmim:2009zm,Adamson:2012gt,Aartsen:2013jza}. In these experiments, in the most common detection channel a neutrino interacts with a nucleus in or around the detector, creating a $\gtrsim$ MeV outgoing lepton (typically a muon) whose direction is reconstructed to tag a neutrino interaction. Through these types of detection, detailed measurements of atmospheric neutrinos have not only confirmed the basic prediction of neutrino production, but also have been important in identifying new physics. For example, Super-Kamiokande measured the ratio of muon to electron type events, and established vacuum-induced $\nu_\mu$ to $\nu_\tau$ transitions as the solution to the zenith angle dependence of this ratio~\cite{Fukuda:1998mi}. 

\par While the atmospheric neutrino flux for energies $\gtrsim 1$ GeV has been well studied by the aforementioned experiments, the low-energy flux of atmospheric neutrinos, $\lesssim 100$ MeV, is difficult to both theoretically model~\cite{Battistoni:2005pd} and to measure. Though the energy spectrum of neutrinos produced corresponds to that of muon and pion decay at rest, the absolute normalization of the flux is less well constrained due to the uncertainties that arise from several physical processes. For example, the cosmic ray flux at the top of the Earth's atmosphere differs from the cosmic ray flux in the interstellar medium. One reason for this is because of the solar wind which decelerates cosmic rays that enter into the heliosphere. A second reason is due to the geomagnetic field, which induces a cut-off in the low-energy cosmic ray spectrum. Detailed modeling of both of these effects implies that for energies $\lesssim 100$ MeV, the uncertainty on the predicted atmospheric neutrino flux is approximately 20\%~\cite{Honda:2011nf,Zhuang:2021rsg}. Due in particular to the cutoff in the rigidity of cosmic rays induced by the Earth's geomagnetic field at low energies, the atmospheric neutrino flux is larger for detectors that are nearer to the poles~\cite{Honda:2011nf,Zhuang:2021rsg}. 

\par The atmospheric neutrino flux is nearly isotropic, with small predicted deviations. Over all energies, the atmospheric neutrino flux peaks near the horizon, at zenith angle $\cos \theta \simeq 0$. At high energies, the flux is very nearly symmetric about $\cos \theta \simeq 0$, as at these energies the cosmic ray particles are more energetic than the rigidity cutoff. At low energies, the flux becomes asymmetric, as the flux of downward-going ($\cos \theta = 1$) neutrinos is lower than the flux of upward-going neutrinos ($\cos \theta = -1$). There is also a time variation in the neutrino flux with a period of $\sim 11$ years due to the modulation of the primary proton cosmic-ray spectrum by the Solar wind~\cite{Zhuang:2021rsg}. This effect is most prominent for detectors at high latitudes, such as SURF or SNOlab. 

\par The predicted nuclear recoil energy distribution from atmospheric neutrinos using the CE$\nu$NS detection channel is shown in the left panel of Figure~\ref{fig:nuflux}. The event rates indicate that a detector exposure of $\sim 20$ ton-yr will be required to begin to be sensitive to atmospheric neutrinos. Because the high energies of the nuclear recoils, the effect of the nuclear form factor becomes important; in particular variations from the standard helm form factor have a significant impact on the predicted rate. Because the form factor is sensitive to the neutron distribution in the nucleus, and there are no laboratory measurements that have been made of this distribution in a nucleus like Xe or Ar, this form factor will likely remain a significant systematic uncertainty in determining the event rate. 

\begin{figure}[!htb]
\centering 
\includegraphics[width=3.0in]{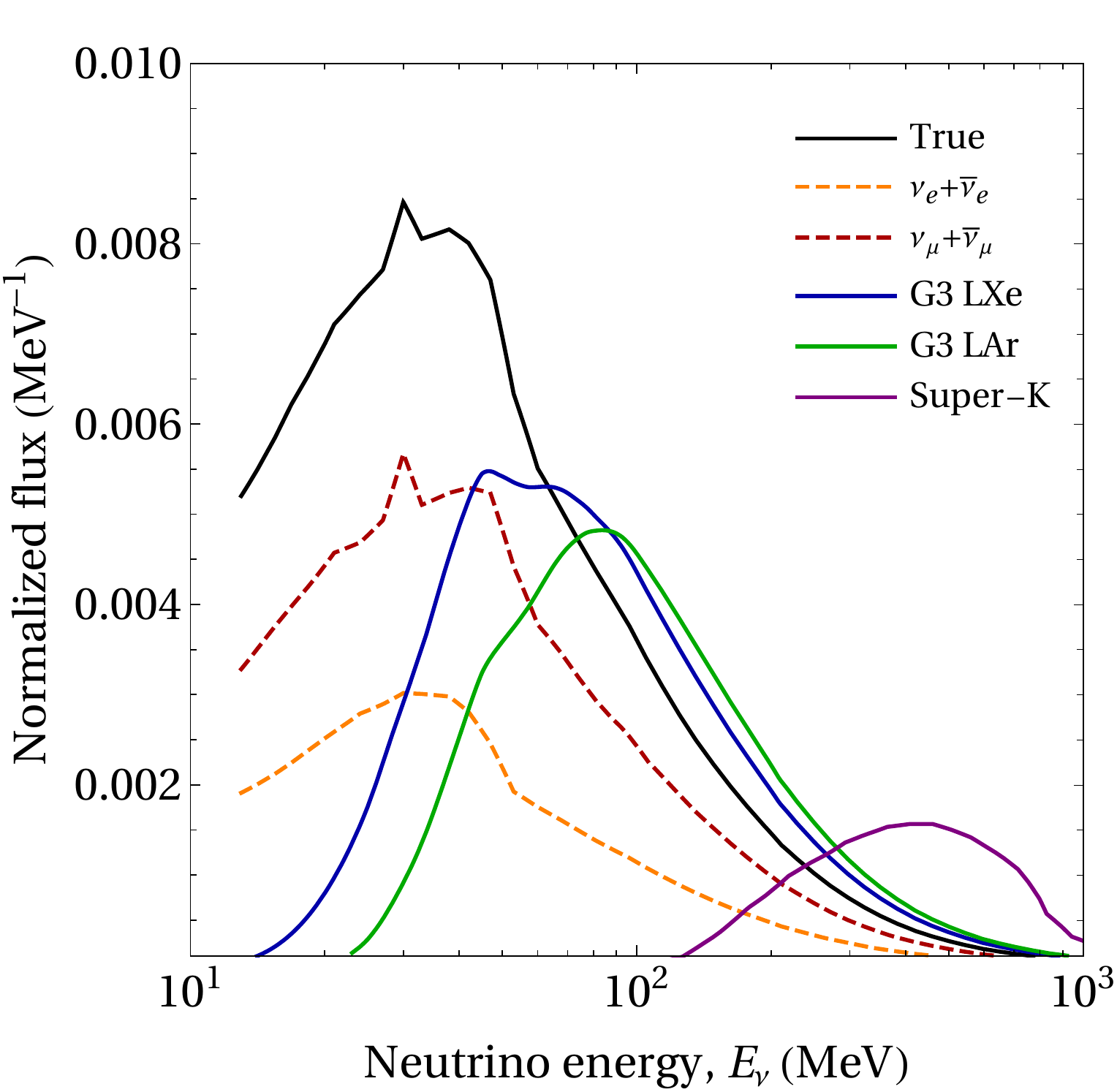}
\caption{The differential fluxes of atmospheric neutrinos that are accessible by various experiments, normalized to unity. The electron and muon-flavored fluxes are indicated with the dashed curves, and the solid black curve is the total atmospheric neutrino flux, summed over all flavors. The features in the neutrino fluxes result from pion and muon decay at rest. Future dark matter experiments will access an atmospheric neutrino energy range that is not accessible to Super-K. From Ref.~\cite{Newstead:2020fie} 
\label{fig:atmdistributions}}
\end{figure}

From the kinematic limits we can find that a detector sensitive to nuclear recoils in the energy range $\sim 1-50$~keV will be sensitive to neutrinos in the energy range $\sim 40-60$~MeV. More precisely, we can asses the range of energies of atmospheric neutrinos a given detector is sensitive to by integrating over $E_R$ above a specified threshold. The result of this integration as a function of $E_\nu$ is given in Fig~\ref{fig:atmdistributions}, indicating the neutrino energy range that a xenon and argon detector with $E_R \geqslant 3$~keV and $E_R \geqslant 25$~keV (respectively) would be sensitive to. For comparison we also show the lowest energy channel (sub-GeV single-ring electron-like events) that Super-Kamiokande was sensitive to in their atmospheric neutrino analysis~\cite{Richard:2015aua}. As indicated, Super-Kamiokande is sensitive to neutrinos $\gtrsim 100$~MeV for their fully-contained electron-like events. 

\section{Beyond the Standard Model physics} 
\label{sec:bsm}
Using the terrestrial and astrophysical sources discussed above, CE$\nu$NS has proven to be a valuable probe of BSM physics. Two specific BSM physics scenarios that can be probed are those that include sterile neutrinos or NSI. This section reviews CE$\nu$NS probes of these topics, highlighting current experimental constraints and those expected in the future. 

\subsection{Sterile neutrinos}

New gauge singlet fermions are a minimal extension of the SM. As long as there are no new symmetries forbidding such a term, gauge and Lorentz invariance allow one to write down the following term in the Lagrangian
\begin{equation}
    \mathcal{L}_{{\rm sterile}} \supset y NHL,
    \label{eq:sterile}
\end{equation}
where $y$ is a Yukawa coupling, $H$ and $L$ are the Higgs and lepton doublets, while $N$ is the new gauge singlet fermion, more commonly referred to as a ``sterile neutrino'' or ``neutral heavy lepton.'' Notice that after electroweak symmetry breaking, $NHL \longrightarrow \langle H \rangle N \nu$, the active neutrinos and the sterile neutrinos mass mix. Thus the existence of such BSM states can explain the observation of neutrino masses. 

Depending on model-dependent details the coupling there may also be a Majorana mass term for $N$. Unlike the charged fermions of the SM, there are no good theoretical arguments (i.e.\ anomaly cancellation) constraining the number of these sterile neutrinos. Moreover, there is no firm theoretical guidance as to what mass scale to associate with these states, as reasonable models have been constructed with the sterile neutrinos ranging from sub-eV to beyond the GUT scale.


Sterile neutrinos as described above can be searched for in a number of experiments. Most searches for sterile neutrinos fall into one of two categories: (1) modified oscillations, or (2) direct production. The second category exploits the fact that sterile neutrinos inherit a portion of the weak interaction via their mixing with the active neutrinos. This allows for their production in meson decays or neutrino scattering, and typically makes them unstable. 

In searching for signatures of sterile neutrinos, experimental data are most easily interpreted within the two-neutrino picture. This picture approximates that the mass-splittings between active neutrino flavors is zero, and that the oscillations are driven by the much larger mass splitting between the active and sterile states. Depending on the source and the detector, experiments are able to probe neutrino appearance ($\nu_\mu \rightarrow \nu_e$, $\bar \nu_\mu \rightarrow \bar \nu_e$), or disappearance ($\nu_\mu \rightarrow \nu_\mu$, $\bar \nu_e \rightarrow \bar \nu_e$). 

Data from several experiments are consistent with a sterile neutrino interpretation with a mass splitting $\gtrsim 1$ eV$^2$. Accelerator appearance experiments LSND~\cite{Aguilar:2001ty} and MiniBooNE~\cite{Aguilar-Arevalo:2018gpe,MiniBooNE:2020pnu} have identified an excess of events in $\nu_\mu \rightarrow \nu_e$ oscillation data. The radioactive source experiments of the GALLEX and SAGE Solar neutrino detectors have found indications of a deficit of electron neutrinos~\cite{Giunti:2006bj,Giunti:2010zu}. Independently, very short baseline neutrino experiments with distances of $<$ 100 m find evidence for a deficit of electron anti-neutrinos~\cite{Mention:2011rk}, however, more recent re-evaluations of these experiments with updated antineutrino flux predictions yield no strong preference for this deficit~\cite{Giunti:2021kab,Berryman:2021yan}. 
There are currently no disappearance experiments utilizing the $\nu_\mu \rightarrow \nu_\mu$ channel that are consistent with a sterile neutrino interpretation, with the exception of a ${\sim}90\%$ CL preference from IceCube~\cite{IceCube:2020tka}. Summaries of sterile neutrino searches with joint fits can be found in Refs.~\cite{Dentler:2018sju,Diaz:2019fwt,Boser:2019rta}, and global analyses of reactor data are given in Ref.~\cite{Berryman:2020agd,Berryman:2021yan}. More recently, the MicroBooNE experiment has begun searching for anomalous $\nu_e$ appearance in a $\nu_\mu$ beam to test this scenario~\cite{MicroBooNE:2021rmx,MicroBooNE:2021sne,MicroBooNE:2021jwr,MicroBooNE:2021nxr}. These results have been reinterpreted in the context of a sterile neutrino search in Ref.~\cite{Arguelles:2021meu}, finding that their null results do not yet rule out the parameter space preferred by MiniBooNE~\cite{MiniBooNE:2022emn} and LSND.

Because of its sensitivity to the total active neutrino flux, CE$\nu$NS experiments are unique in their capability to search for sterile neutrinos. At a fixed distance baseline, the signature of sterile neutrinos would be a depletion of the flux relative to that predicted by the SM. This requires a precise understanding of the systematic uncertainties on the neutrino flux from the source. Sterile neutrinos may be identified by comparing the energy spectrum of nuclear recoil events at different distance baselines. This technique is independent of the systematic uncertainties associated with the source flux normalization, though does require detectors with sufficient energy resolution.   

\begin{figure}
    \centering
    \includegraphics[width=0.5\textwidth]{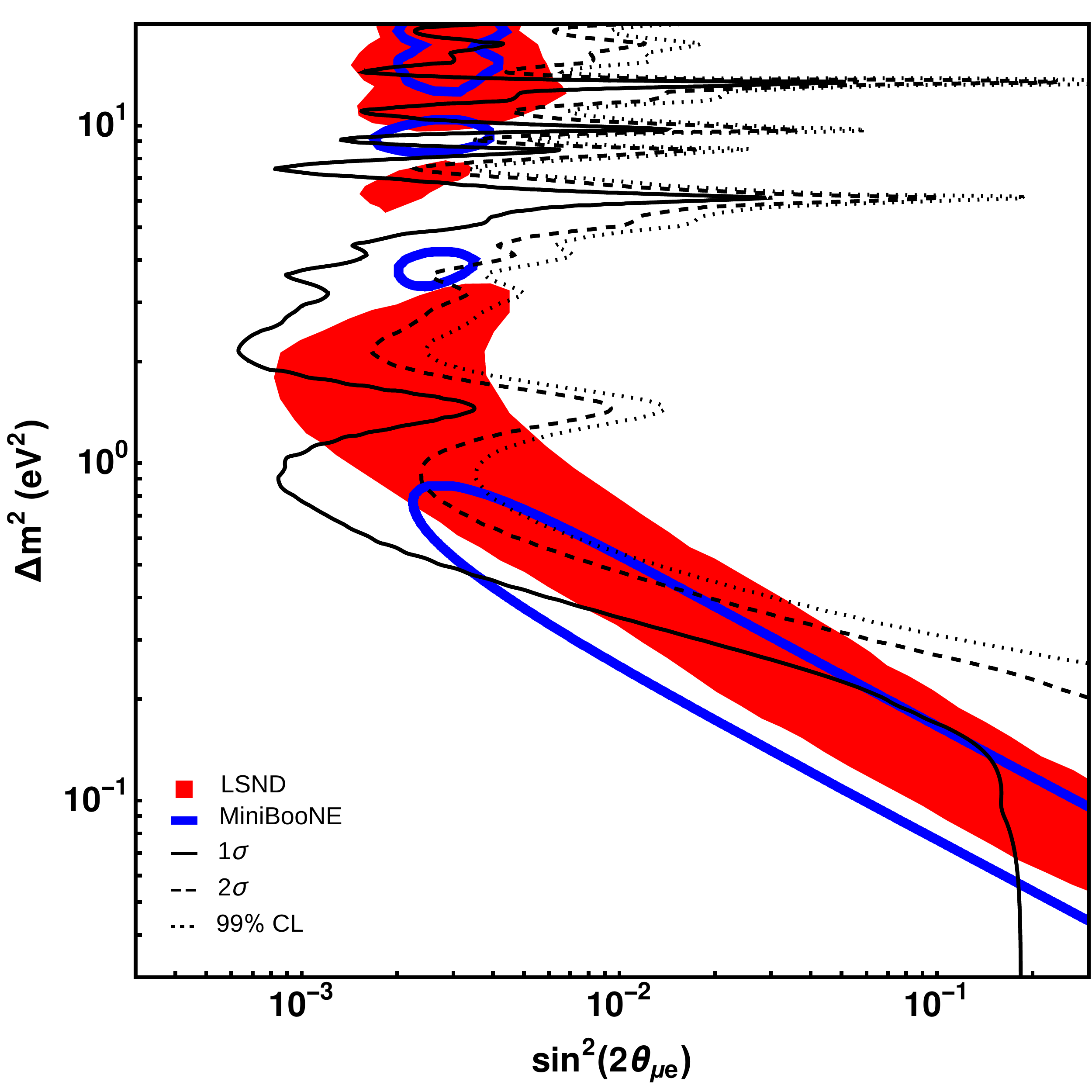}
    \caption{Constraints on the sterile neutrino mass splitting and mixing angle assuming a 3 year exposure from COHERENT, with distance baselines of 20 and 40 m. Closed contours show the allowed LSND and MiniBooNE regions (figure taken from Ref.~\cite{Blanco:2019vyp}).
    \label{fig:sterile}}
\end{figure}

CE$\nu$NS detectors at stopped-pion sources may be purposed for sterile neutrino searches. In particular, the sensitivity to sterile neutrinos is maximized when deploying multiple detectors at different distance baselines in the range $\sim 20$--$40$ m~\cite{Anderson:2012pn}. This configuration can probe parameter space that is consistent with the $\sim$ eV mass-scale hinted at by LSND and MiniBooNE, thereby providing an independent test of sterile neutrino parameter space~\cite{Anderson:2012pn,Blanco:2019vyp}. Figure~\ref{fig:sterile} shows projected constraints for a stopped-pion experiment, assuming two different baselines. This shows that CE$\nu$NS is able to provide a strong independent probe over nearly the entire mass splitting and mixing regime. 

CE$\nu$NS detectors deployed with a reactor sources may also be utilized to search for signatures of sterile neutrinos. As discussed in detail below, several reactor-based experiments, using baselines anywhere between $\sim 2$--$20$ m, are now being designed to detect CE$\nu$NS. These experiments will provide the capability to test the sterile neutrino interpretations of existing data sets. In contrast to the case of the stopped-pion source, reactor-based experiments are only sensitive to the mixing of the electron neutrino with the sterile state. 
Finally, CE$\nu$NS has been proposed as a new probe
of lepton unitarity and sterile neutrinos~\citep{Miranda:2020syh,Denton:2021mso}.

\subsection{Additional neutrino interactions}
\label{sec:NSI}

In this section we review the main phenomenological implications derived from the inclusion of additional operators leading to modifications of vector- and axial-vector interactions in the SM with a non-trivial flavor structure, as well as from the addition of new operators with a different Lorentz structure (e.g., scalar or tensor operators). We also review the phenomenological implications of neutrino electromagnetic properties in the context of CE$\nu$NS.


\subsubsection{Non-Standard Interactions: Vector and axial-vector interactions}
\label{sec:vector_axial_vector}


A convenient way to parameterize possible new physics effects in low-energy observables is through the addition to the SM Lagrangian of effective higher-dimensional operators. In the context of neutrino physics, the term Non-Standard Interactions (NSI) usually refers to the inclusion of four-fermion ($d=6$) operators leading to modifications of the Wilson coefficients already present in the SM~\cite{Wolfenstein:1977ue} (for recent reviews see, e.g.,  Refs.~\cite{Farzan:2017xzy,Dev:2019anc}, or Refs.~\cite{Altmannshofer:2018xyo,Hoferichter:2020osn} for the operator basis relevant for CE$\nu$NS).  
In the context of CE$\nu$NS, NSI of interest are those affecting Neutral-Current (NC) processes involving quarks and neutrinos, that is:
\begin{equation} 
\label{eq:NSI}
\mathcal{L}_{\rm NSI} \supset -2\sqrt{2} G_F \sum_{\alpha,\beta}\sum_{P,q}  \epsilon^{q,P}_{\alpha \beta}( \bar{\nu}_\alpha \gamma^\mu P_L \nu_\beta)( \bar{q} \gamma_\mu P q),
\end{equation} 
where $q \in \{u,d\}$ and $P\equiv P_L,P_R$, and hermiticity requires that $\epsilon_{\alpha\beta}^{q,P} = (\epsilon_{\beta\alpha}^{q,P})^*$, with $\alpha,\beta = (e,\mu,\tau)$. In the literature these are often rearranged to make explicit their effect on the SM vector and axial-vector Fermi operators at low energies, as:
\begin{equation}
\label{NSI_EFT}
    \mathcal{L}_{\rm NSI} \supset -\sqrt{2} G_{F} \sum_{\alpha,\beta} \sum_q \Big[\epsilon_{\alpha \beta}^{qV}\bar{\nu}_{\alpha}\gamma_{\mu}P_L \nu_{\beta} \,\bar{q}\gamma^{\mu} q-\epsilon_{\alpha \beta}^{qA}\bar{\nu}_{\alpha}\gamma_{\mu}P_L \nu_{\beta} \,\bar{q}\gamma^{\mu}\gamma_5 q\Big]\,,
\end{equation}
where we have defined $\epsilon_{\alpha \beta}^{qV}=\epsilon_{\alpha \beta}^{qL}+\epsilon_{\alpha \beta}^{qR}$, $\epsilon_{\alpha \beta}^{qA}=\epsilon_{\alpha \beta}^{qL}-\epsilon_{\alpha \beta}^{qR}$, leading to
\begin{equation}
 C_q^V\to C_q^V\big|_\text{SM}+\epsilon^{qV}\,,\qquad 
 C_q^A\to C_q^A\big|_\text{SM}-\epsilon^{qA}\,,
\end{equation}
for the Wilson coefficients defined in Eq.~\eqref{Lagr_SM}. 

CE$\nu$NS experiments are primarily sensitive to vector NSI (though axial-vector interactions may be present in neutral current NSI, they only become significant compared to vector interactions for relatively light nuclei). Specifically, operators in Eq.~\eqref{NSI_EFT} modify the weak charge as~\cite{Barranco:2005yy}
\begin{align}
\label{eq:QWnsi}
    \Big(Q_{\text{w}}^{\nu_\alpha}\Big)^2&=
    \left[Z\big(Q_\text{w}^{\nu_\alpha,p} + 
    2\epsilon^{p,V}_{\alpha\alpha}\big) + 
    N\big(Q_\text{w}^{\nu_\alpha,n}+
    2\epsilon^{n,V}_{\alpha \alpha}\big)\right]^2 +4\sum_{\beta\neq\alpha}\left[Z \epsilon^{p,V}_{\alpha\beta} +N\epsilon^{n,V}_{\alpha\beta}\right]^2\,,
\end{align}
with 
\begin{equation}
    \epsilon_{\alpha\beta}^{p,V} \equiv 2 \epsilon_{\alpha\beta}^{u,V} + \epsilon_{\alpha\beta}^{d,V} \, , \qquad
    \epsilon_{\alpha\beta}^{n,V} \equiv \epsilon_{\alpha\beta}^{u,V} + 2 \epsilon_{\alpha\beta}^{d,V} \, ,
\end{equation} 
where we have assumed real NSI, and radiative corrections for the weak charges of the proton and the neutron have been included, so they explicitly depend on the neutrino flavor index (see Sec.~\ref{sec:radiative}). Note also that the weak form factor changes in presence of NSI~\cite{Hoferichter:2020osn}. 

\begin{figure}
    \centering
    \includegraphics[width=0.5\textwidth]{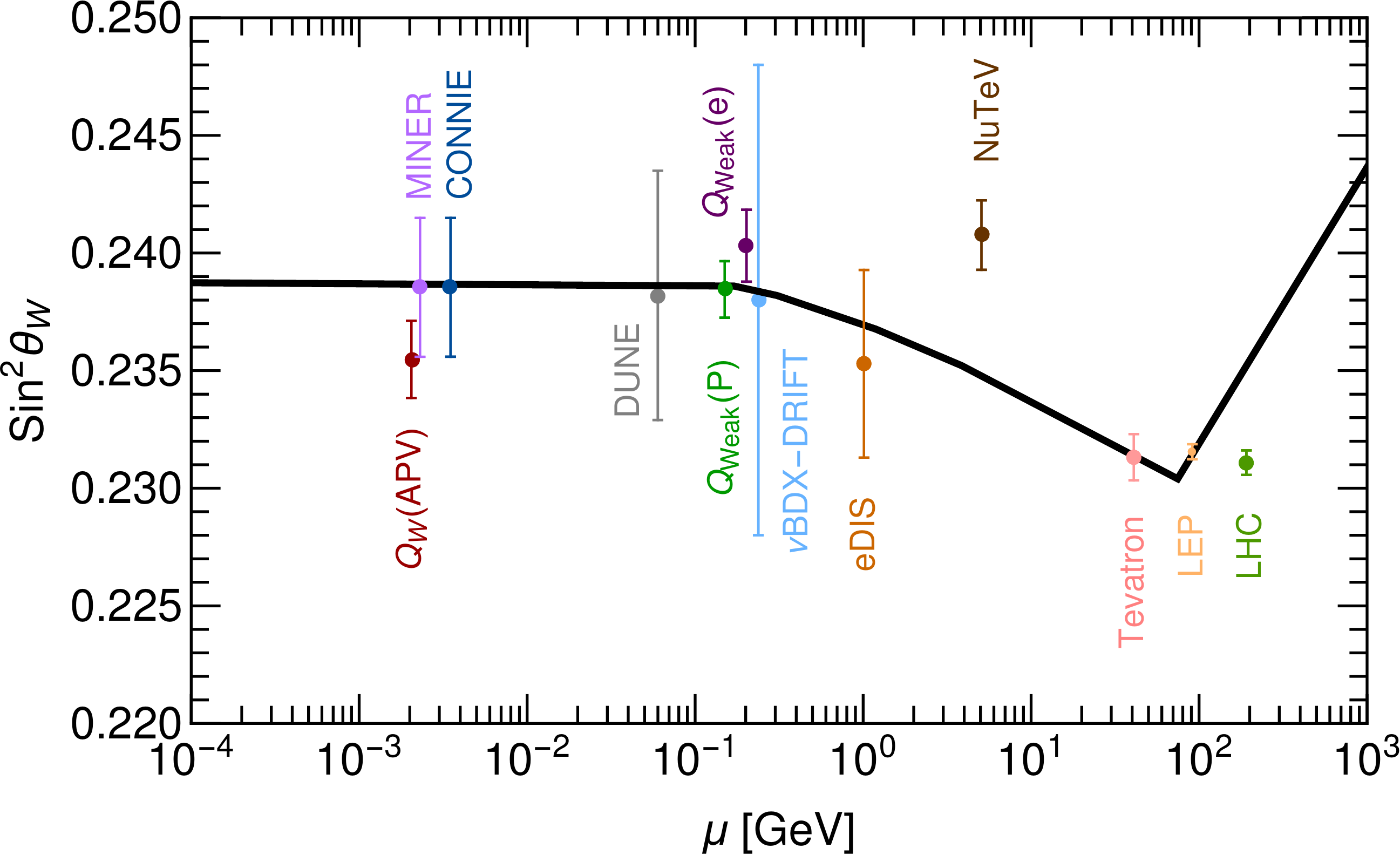}
    \caption{Sensitivity of CE$\nu$NS experiments to $\sin^2\theta_W$ compared to other SM precision tests (figure adapted from Ref.~\cite{AristizabalSierra:2021uob}).
    \label{fig:s2w}}
\end{figure}

From a phenomenological perspective, NSI may show up as a modification of the weak mixing angle that is experimentally measured. As described in Sec.~\ref{sec:SM}, the normalization of the CE$\nu$NS cross section depends on the weak charge of the nucleus (with possible corrections from axial-vector interactions in case the target nucleus carries spin), which in the SM depends on $\sin^2\theta_W$. A possible deviation of the measured weak charge from its SM value (including the appropriate radiative corrections discussed in Sec.~\ref{sec:radiative}) could then be interpreted as a signal from new physics. This has been made evident in our notation in Eq.~\eqref{eq:QWnsi}: new physics models inducing NSI with protons will directly affect the extraction of $\sin^2\theta_W$. Formulating constraints in terms of $\sin^2\theta_W$ thus defines a minimal scheme to compare the sensitivity of CE$\nu$NS experiments to other SM precision tests, including parity-violating electron scattering and electroweak precision observables~\cite{Canas:2018rng,Fernandez-Moroni:2020yyl,AristizabalSierra:2021uob,Crivellin:2021bkd,Cadeddu:2021ijh,Cadeddu:2019eta,Cadeddu:2020lky,Cadeddu:2018izq,Sierra:2022ryd}. The projected sensitivity is shown in Fig.~\ref{fig:s2w} for a selected set of CE$\nu$NS experiments. 

\begin{figure}
    \centering
    \includegraphics[width=0.49\textwidth]{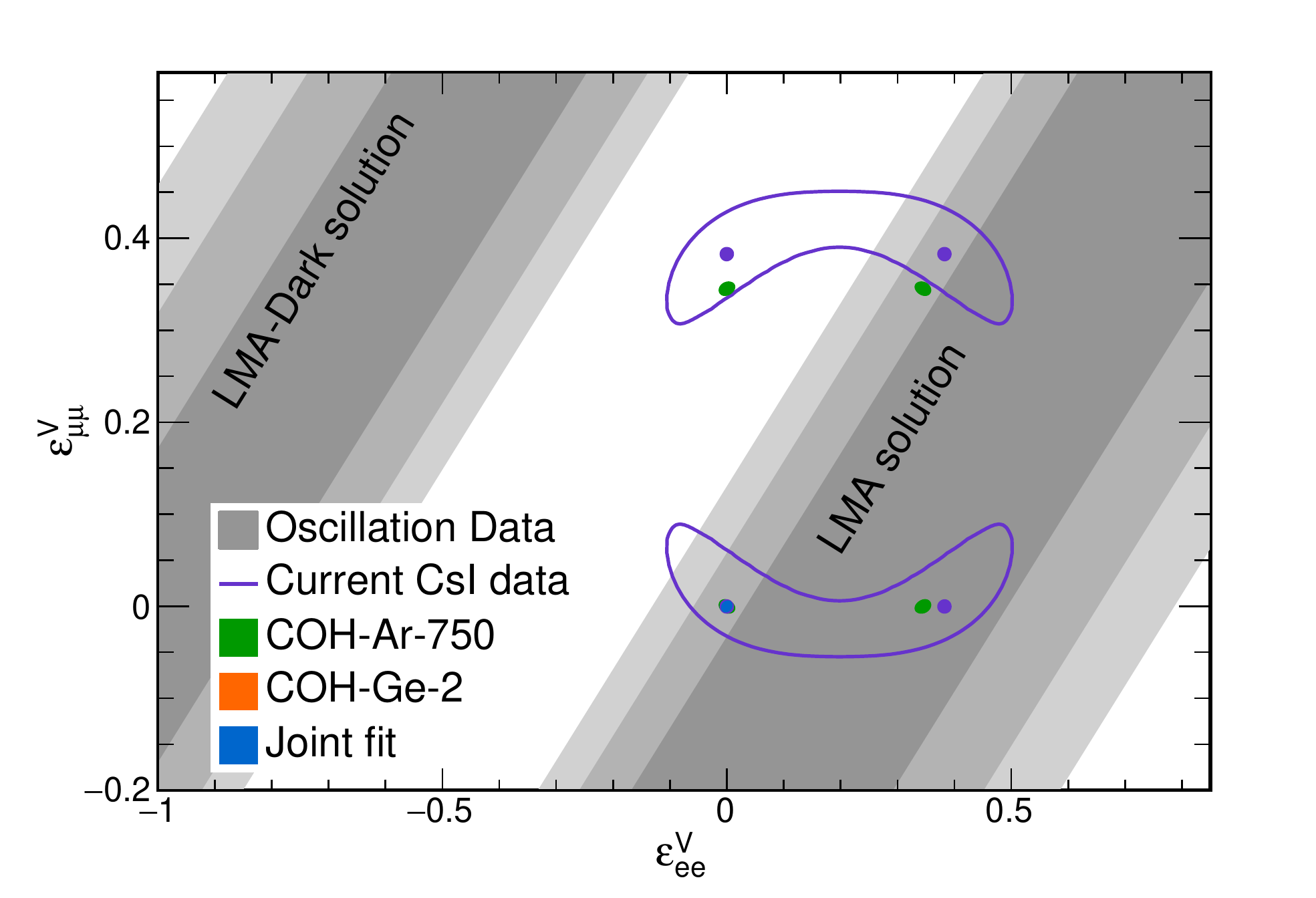}
    \includegraphics[width=0.49\textwidth]{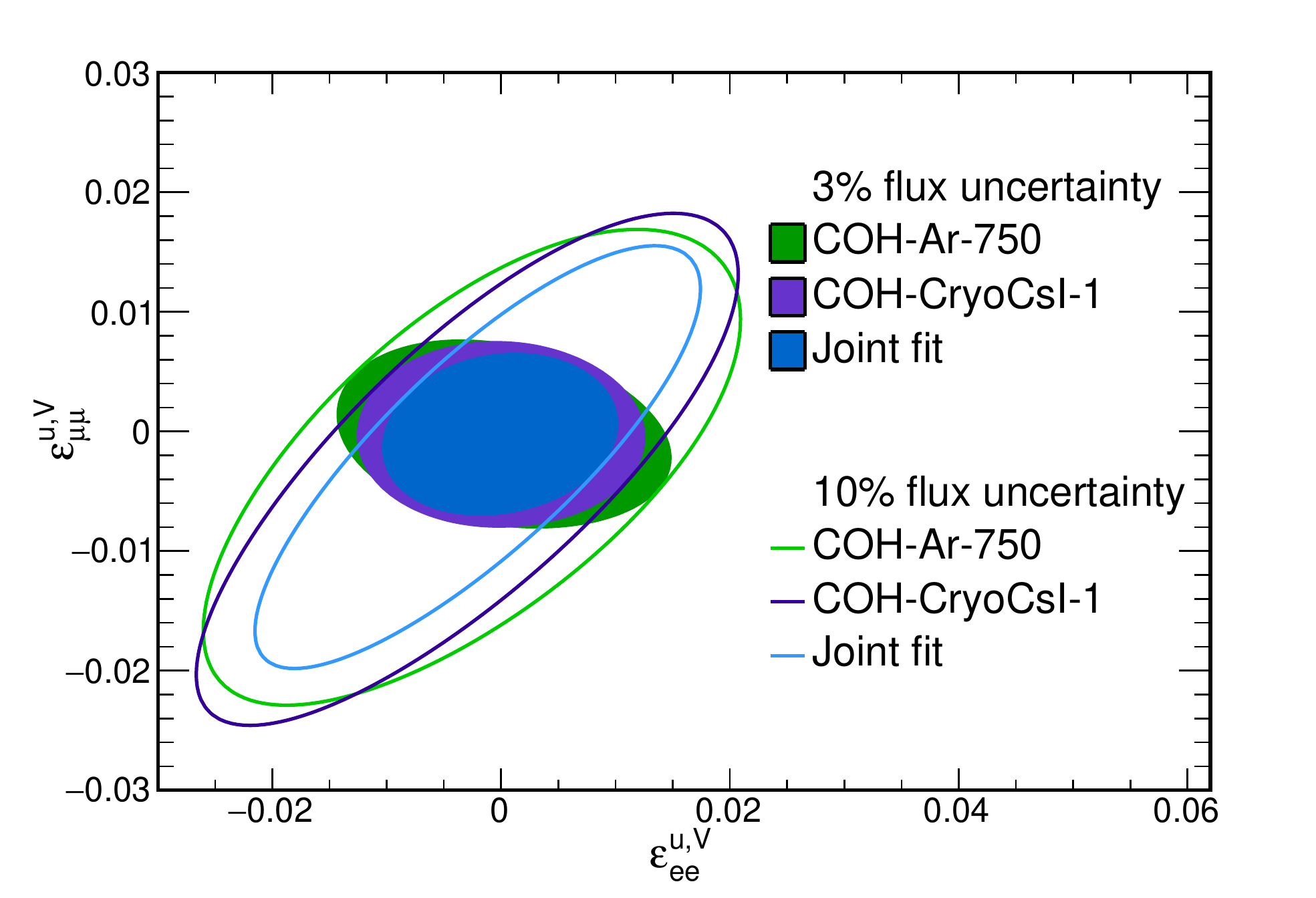}
    \caption{COHERENT sensitivity (90$\%$) to disfavor the LMA-Dark solution at the SNS first target station with current data overlayed. The right panel shows parameter space allowed by a joint fit of COHERENT CEvNS measurements on different nuclei.
    \label{fig:LMAMSW}}
\end{figure}

In a more general context, NSI affect the matter potential that neutrinos feel as they propagate in a medium and therefore may lead to significant modifications of the neutrino oscillation probabilities. Global fits to oscillation data have been performed for NSI with quarks~\cite{Esteban:2018ppq,Gonzalez-Garcia:2011vlg, Gonzalez-Garcia:2013usa} and set relatively strong bounds on the size of the off-diagonal NSI coefficients; however, oscillation experiments are only sensitive to differences between the diagonal NSI parameters. Neutrino scattering (and, in particular, CE$\nu$NS), on the other hand, is sensitive to the diagonal NSI parameters individually and thus provides complementary information to that from oscillations~\cite{Coloma:2017ncl}. For recent works deriving bounds on NSI using current CE$\nu$NS data, see, e.g., Refs.~\cite{Coloma:2017ncl,Coloma:2019mbs,Coloma:2022avw,Liao:2022hno,Liao:2017uzy,Giunti:2019xpr,Papoulias:2017qdn,Khan:2021wzy,Denton:2020hop}; for future prospects see, e.g., Refs.~\cite{Dent:2017mpr,Billard:2018jnl,Dutta:2020che,Baxter:2019mcx,Shoemaker:2021hvm}. Besides improving the overall sensitivity to NSI parameters, the combination of oscillation and scattering data also disfavors the LMA-Dark solution~\cite{Miranda:2004nb,Escrihuela:2009up}. This is relevant for the determination of the neutrino mass ordering at current and future oscillation experiments~\cite{Gonzalez-Garcia:2011vlg,Gonzalez-Garcia:2013usa,Bakhti:2014pva,Coloma:2016gei}, as discussed in more detail in Ref.~\cite{neutrino_BSM_WP}. Examples for CE$\nu$NS constraints on the NSI parameters are shown in Fig.~\ref{fig:LMAMSW}.

\subsubsection{New neutrino interactions in $U(1)'$ models}
\label{sec:light_mediators}

At this point it is worth noting that the interactions included in Eq.~\eqref{eq:NSI} are not gauge invariant. If these operators are obtained from a new theory at high energies, gauge invariance generically implies the simultaneous generation of similar operators involving charged leptons, for which tight experimental constraints exist~\cite{Gavela:2008ra,Antusch:2008tz}. Unless fined-tuned cancellations are invoked~\cite{Gavela:2008ra}, this makes it hard to build a model that leads to sizable NSI effects in neutrino experiments (either in oscillations or in CE$\nu$NS experiments). A possibility to avoid the tight bounds from charged-lepton flavor violating observables is to consider that the new physics may be weakly coupled to the SM, via new mediators with masses well \emph{below} the EW scale (see the related discussion in Ref.~\cite{neutrino_BSM_WP}). A well-motivated example is to extend the SM with an extra $U(1)'$ symmetry, which may serve as a vector portal to the dark sector. The new mediator associated to this symmetry is usually referred to as a $Z'$.

Matching a concrete new physics model onto the EFT at low energies, a dependence on the mediator mass arises, since the amplitude of CE$\nu$NS depends on its propagator. In this case, the weak charge of the nucleus gets modified in a similar way as in the NSI case, simply replacing
\begin{eqnarray}
\epsilon_{\alpha \alpha}^{qV} \to \frac{(c_\alpha^\prime g_{Z^\prime})^2}{\sqrt{2} G_F (q^2 + M_{Z^\prime}^2)}\,,
\label{eq:NSIlight} \quad 
\epsilon_{\alpha \beta}^{qV} \to 0 \, ,
\end{eqnarray}
where for simplicity we have assumed that the new interaction is lepton-flavor conserving. Here, $c_\alpha^\prime$ refers to the charge of $\nu_\alpha$ under the new interaction, while $g_{Z^\prime}$ refers to the new gauge coupling introduced. From this discussion it follows that, for a weakly-coupled $Z'$, bounds from scattering experiments with momentum transfer above the mediator mass ($q^2 \gg M_{Z'}^2$) will be suppressed since the cross section in this case is proportional to $ \sim g_{Z'}^2/q^2$. CE$\nu$NS measurements become very relevant in this context, since the very low momentum transfers involved allow this process to be sensitive to a wider set of $Z'$ models. Recent works studying bounds on light vector mediators with current CE$\nu$NS data include Refs.~\cite{Coloma:2020gfv,Coloma:2022avw,Corona:2022wlb,Denton:2018xmq, Cadeddu:2020nbr,Dent:2016wcr,Papoulias:2017qdn,Miranda:2020tif,Abdullah:2018ykz,CONUS:2021dwh,CONNIE:2019xid,delaVega:2021mhj,Shoemaker:2017lzs}. Sensitivities at future facilities have been studied for the European Spallation Source~\cite{Bertuzzo:2021opb} and Los Alamos National Laboratory~\cite{Shoemaker:2021hvm}.



\subsubsection{Electromagnetic properties}
\label{sec:electromagnetic}

The fact that neutrinos are massive, as implied by the robust discovery of neutrino oscillations, provides the best motivation for the existence of non-trivial neutrino electromagnetic (EM) properties.
These usually come in the form of a neutrino magnetic moment or a neutrino charge radius.
If sizable enough, they could in principle induce detectable distortions of the expected signal at CE$\nu$NS experiments.
The differential cross section in the presence of a neutrino magnetic moment adds incoherently to the Standard Model cross section due to the required spin-flip, i.e., $
\left( \frac{d \sigma}{dT}\right)_{\mathrm{tot}} = \left( \frac{d \sigma}{dT} \right)_{\mathrm{SM}} 
+ \left( \frac{d \sigma}{dT} 
\right)_{\mathrm{EM}}$, 
where the EM contribution has a characteristic $1/T$ dependence, while its strength is controlled by the size of the effective neutrino magnetic moment $\mu_{\nu}^\text{eff}$, as~\cite{Vogel:1989iv}
\begin{equation}
\left( \frac{d \sigma}{dT} \right)_{\mathrm{EM}}=\frac{\pi \alpha^2 (\mu_{\nu}^\text{eff})^{2}\,Z^{2}}{m_{e}^{2}}\left(\frac{1-T/E_{\nu}}{T}+\frac{T}{4E_\nu^2}\right) F_\text{ch}^{2}(q^{2})\, ,
\label{NMM-cross section}
\end{equation}
with $\alpha$ being the fine-structure constant and $F_\text{ch}$, normalized as $F_\text{ch}(0)=1$, the charge form factor of the nucleus. On the other hand, the impact of the neutrino charge radius, being a helicity-preserving quantity, is taken as a shift on the weak mixing angle according to 
\begin{equation}
\sin^2 \theta_W \rightarrow \sin^2 \theta_W + \frac{\sqrt{2} \pi \alpha}{3 G_F} \langle r_{\nu_\alpha}^2\rangle \, .
\label{eq:rv}
\end{equation}

The recent data from the observation of CE$\nu$NS on CsI and liquid argon (LAr) detectors by the COHERENT experiment have yielded new constraints  on $\mu_{\nu_{\alpha}}^\text{eff}$ and $\langle r_{\nu_\alpha}^2\rangle$ with $\alpha= e, \mu, \tau$~\cite{Cadeddu:2020lky}. The current sensitivities are still relatively weak, i.e., at 90\% C.L.\ they read~\cite{Miranda:2020tif} 
\begin{equation}
\left( \mu_{\nu_e}^\text{eff}, \mu_{\nu_\mu}^\text{eff}, \mu_{\bar{\nu}_\mu}^\text{eff} \right) < (94, 53, 78)~10^{-10}\mu_B\, ,
\end{equation}
and 
\begin{equation}
\begin{aligned}
\langle r_{\nu_e}^2\rangle =& (-64, -41) \, \, \text{and} \, \, (-7, 16)\, , \\  
\langle r_{\nu_\mu}^2\rangle =& (-69, -37) \, \, \text{and} \, \, (-10, 21)\, , \\  
\langle r_{\bar{\nu}_\mu}^2\rangle =& (-60,-43)\, \, \text{and} \, \, (-5, 12)\, , \\  
\end{aligned}
\end{equation}
in units of $~10^{-32}\mathrm{cm^2}$ 
%
%

Analytic expressions for the $\mu_{\nu}^\text{eff}$ can be obtained, starting from the general formula in Ref.~\cite{Grimus:2002vb}, for the different neutrino sources,
e.g., spallation neutron source (SNS), reactors, etc. 
\begin{equation}
\left(\mu_\nu^{\text{eff}} \right)^2 = \tilde{\mathfrak{a}}_{-}^\dagger \tilde{\lambda}^\dagger \tilde{\lambda} \tilde{\mathfrak{a}}_{-} + \tilde{\mathfrak{a}}_{+}^\dagger \tilde{\lambda} \tilde{\lambda}^\dagger \tilde{\mathfrak{a}}_{+} \, ,
\label{eq:TMM-mass}
\end{equation} 
where  the amplitudes of positive and negative helicity states are denoted by the $3$-vectors $\mathfrak{a}_{+}$ and $\mathfrak{a}_{-}$, respectively.   It is therefore convenient to express the transition magnetic moment  (TMM) matrix in the mass basis for Majorana neutrinos in terms of the individual TMMs $\Lambda_i$, as
\begin{equation}
\tilde{\lambda} = \left( \begin{array}{ccc}
0 & \Lambda_3 & - \Lambda_2 \\
- \Lambda_3 &  0 & \Lambda_1 \\
\Lambda_2 & - \Lambda_1 & 0
\end{array} \right) \, .
\label{NMM:matrix}
\end{equation}

On the other hand, the neutrino magnetic moment, observable in a CE$\nu$NS experiment, is in reality an effective parameter that depends on the oscillation parameters and the baseline $L$, as~\cite{Beacom:1999wx}
\begin{equation}
\left(\mu^\text{eff}_{\nu}\right)^2 (L, E_\nu) = \sum_j \Big \vert \sum_i U^\ast_{\alpha i} e^{-i\, \Delta m^2_{ij} L /2 E_\nu} \tilde{\lambda}_{ij} \Big \vert^2 \, ,
\label{NMM-observable}
\end{equation}
where $\tilde{\lambda}_{ij}$ are the elements of the Majorana neutrino TMM matrix in the mass basis.
The individual TMMs can be probed in CE$\nu$NS measurements and their effect has been comprehensively studied in Refs.~\cite{Canas:2015yoa,Miranda:2019wdy}. 
%
%

Reducing the detection threshold is crucial for improving the current sensitivities.
Indeed, reactor-based CE$\nu$NS experiments with sub-keV capabilities should offer more exciting results, complementary to current limits. Very recently, measurements obtained at the Dresden-II reactor experiment~\cite{Colaresi:2022obx} have been interpreted in the context of neutrino magnetic moment searches, providing an upper bound on $\mu_{\nu_e}^{\rm eff}$ below $ \mathcal{O}(\mathrm{few}) \times 10^{-10}\mu_B$, depending on the choice of quenching factor and other technical details of the data analysis~\cite{Coloma:2022avw,Sierra:2022ryd}.
The authors of Refs.~\cite{Miranda:2019wdy,Baxter:2019mcx} have computed the projected sensitivities at the next generation CE$\nu$NS experiments and find that they could compete with the current best upper limits, e.g., from Borexino. 

\subsubsection{Other NSI}
\label{sec:other}

A neutrino dipole moment is only the simplest example of a new class of operators that can be constructed once a chirality flip is admitted, with further interactions including~\cite{Altmannshofer:2018xyo,Hoferichter:2020osn}
\begin{equation}
 \mathcal{L}_{\rm NSI} \supset C_F \bar\nu \sigma^{\mu\nu} P_L\nu F_{\mu\nu}
 +\sum_{q}\Big[C_q^S\bar\nu P_L\nu\, m_q\bar q q+C_q^T\bar\nu \sigma^{\mu\nu}P_L\nu\,\bar q\sigma_{\mu\nu}q\Big]\,,
\end{equation}
where the dipole interaction has been expressed in terms of the Wilson coefficient $C_F$. The cross section for the dipole and scalar operators can be written as 
\begin{equation}
\label{dipole_scalar}
\frac{d \sigma}{d T}\bigg|_\text{dipole+scalar}=\frac{M^2 T}{4\pi E_\nu^2}
\bigg|F_S(q^2)+\frac{2E_\nu-T}{M T}Ze C_F F_\text{ch}(q^2)\bigg|^2\,,
\end{equation}
generalizing Eq.~\eqref{NMM-cross section} and again to be added incoherently to the SM cross section. While the dipole part is determined by the charge form factor of the nucleus, the scalar operator leads to a new form factor $F_S(q^2)$~\cite{Hoferichter:2020osn}. Finally, a tensor operator induces NSI similar to the axial-vector contributions
, but the required form factors differ due to the required multipole decomposition~\cite{Hoferichter:2020osn}.

\section{Experimental efforts} 
\label{sec:experiments} 
Many on-going and planned experiments around the world are dedicated to the detection of CE$\nu$NS. This section reviews these experimental efforts. 

\begin{figure}[!htb]
\centering
\includegraphics[width=4.0in,angle=0]{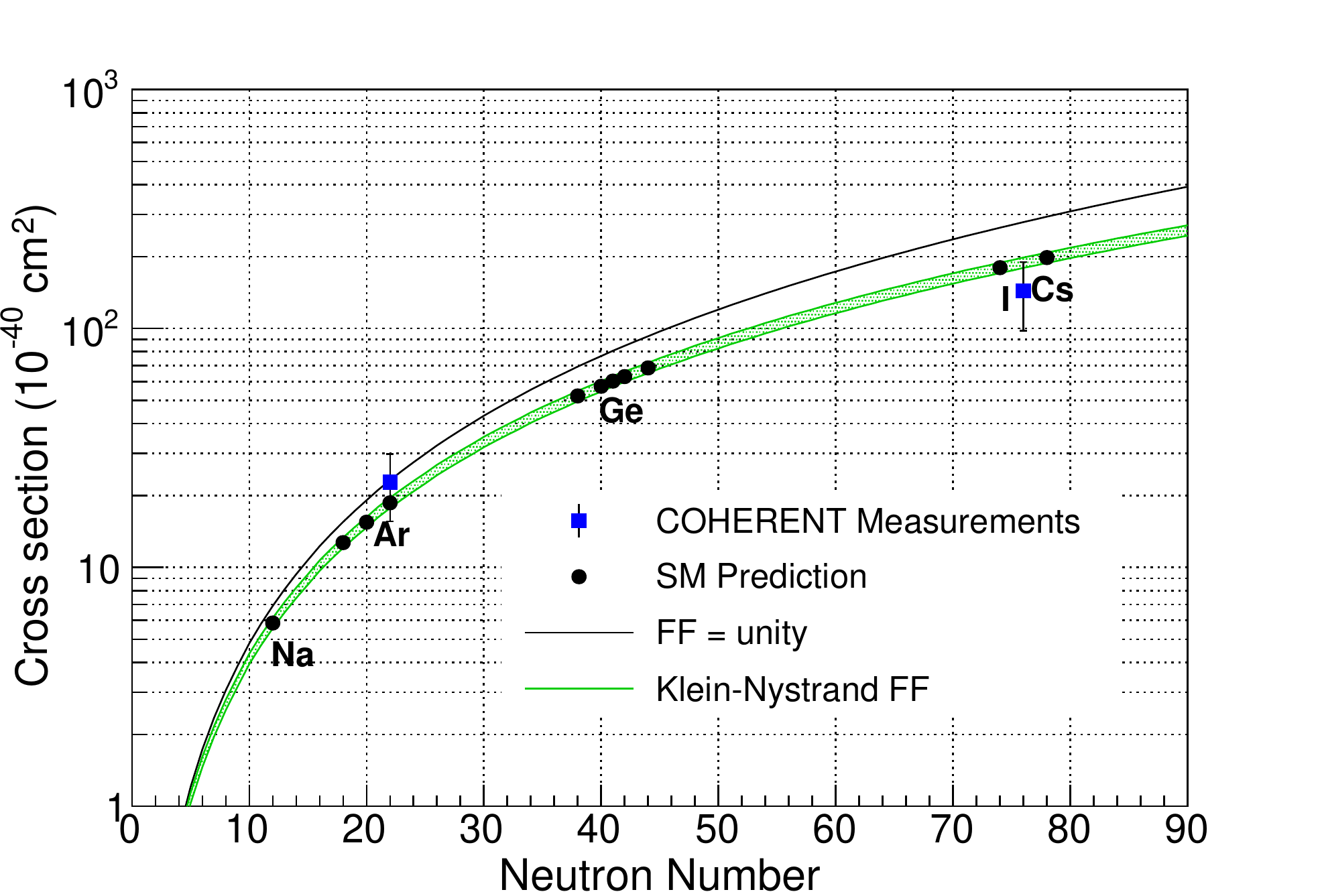}
\caption{Flux-averaged CE$\nu$NS cross section as a function of neutron number for SNS flux, showing existing COHERENT measurements on CsI and Ar, as well as future targets for COHERENT. Figure from ~\cite{Barbeau:2021exu}.} 
\label{fig:kate}
\end{figure}

\subsection{Stopped-pion beams}
As described in Section~\ref{sec:terrestrial}, stopped-pion sources are a source of $\nu_e$, $\nu_\mu$, and $\bar{\nu}_\mu$. The low-energy neutrinos that are produced, $\sim 10-50$ MeV, are a source for CE$\nu$NS. The following experiments are either in data-taking mode, or are being planned, at different stopped-pion sites around the world. The general flux-averaged cross section for a stopped-pion flux is shown in Figure~\ref{fig:kate}. 

\bigskip
\noindent 
\underline{SNS}: The COHERENT collaboration has reported the first detection of coherent neutrino-nucleus elastic scattering (CE$\nu$NS)~\cite{Akimov:2017ade}. COHERENT utilizes the Spallation Neutrino Source (SNS) with a stopped-pion beam, which produces a well-known neutrino spectrum from pion and muon decay at rest. Muon neutrinos, $\nu_\mu$, arrive from prompt charged pion decay, while $\bar{\nu}_\mu$ and $\nu_e$ are produced from the delayed muon decay. With an exposure of 14.6-308 kg-days, the COHERENT collaboration identified nuclear recoil events from CE$\nu$NS which is well in excess of the expected background events for this exposure. 

\par
\begin{figure}[!htb]
\centering
\includegraphics[width=2.30in,angle=0]{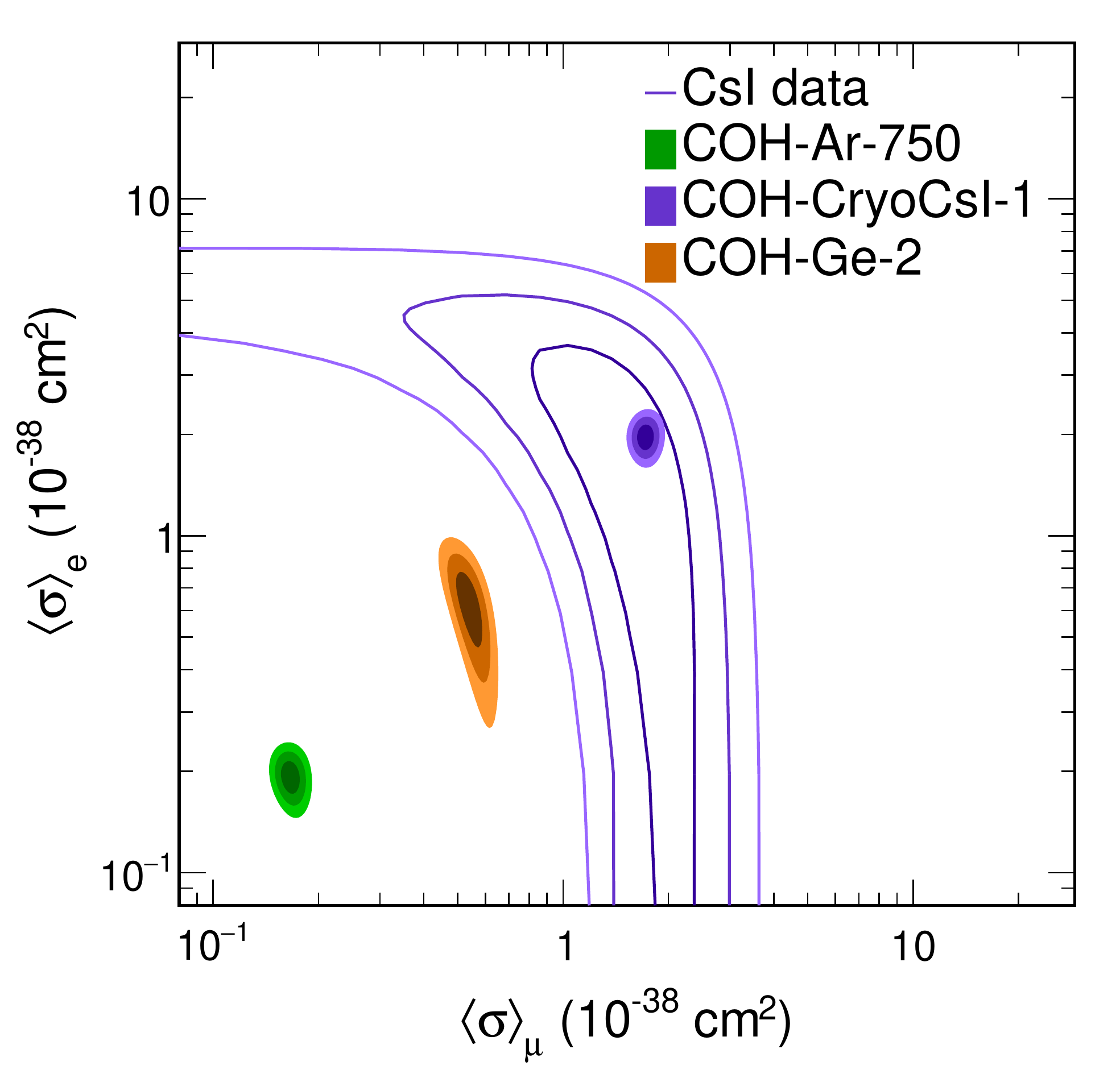}
\includegraphics[width=2.15in,angle=0]{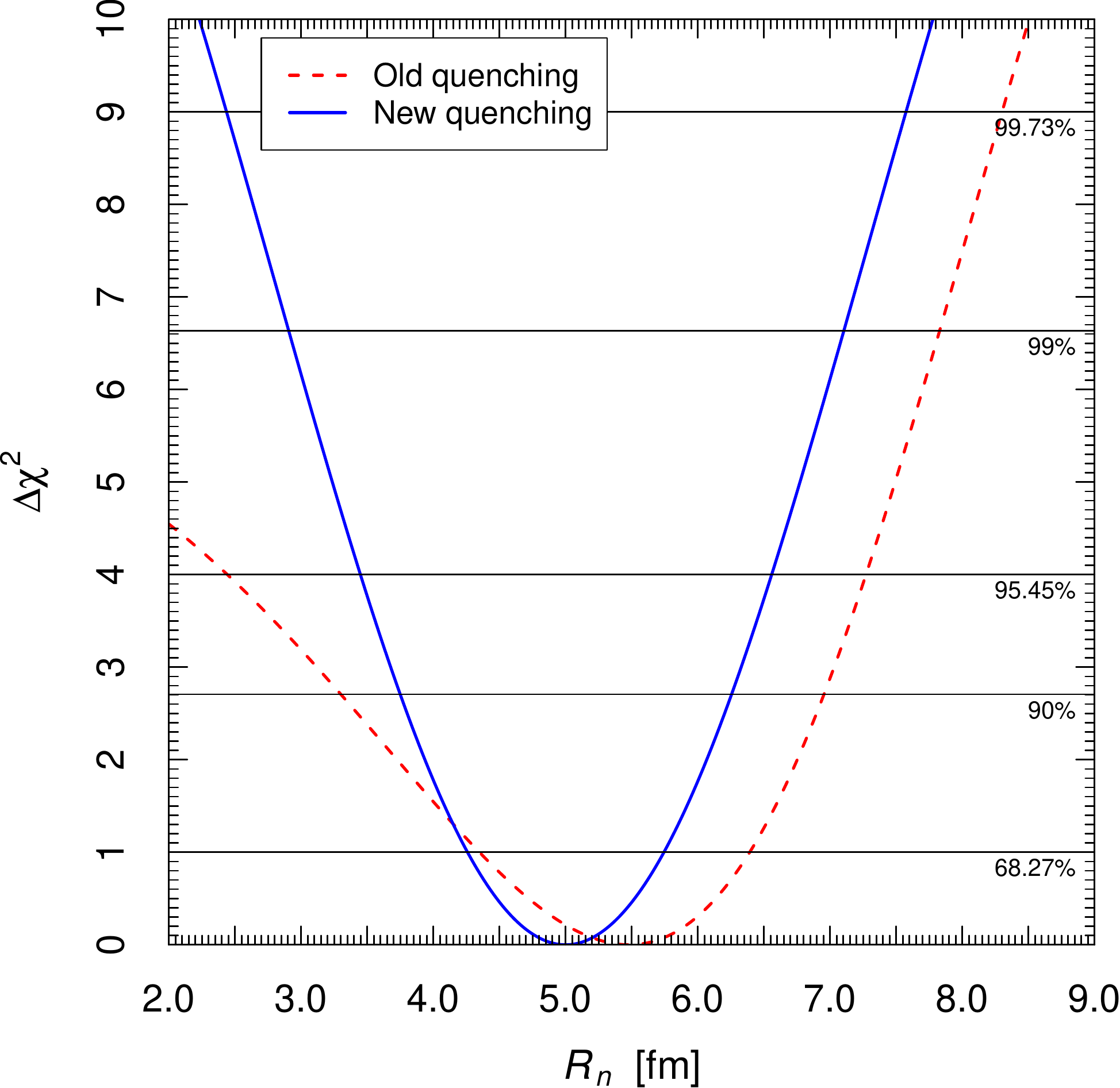}
\caption{The flux-averaged CE$\nu$NS cross section for $\nu_e$ and $\nu_\mu$ flavors.  Current data is compared to future measurements on Ar, Ge, and CsI from COHERENT at the SNS first target station. Right: Measurement of the RMS radius of the neutron distribution for two models of the quenching factor. Figure from Ref.~\cite{Cadeddu:2017etk,Cadeddu:2019eta}.} 
\label{fig:flavors}
\end{figure}

 There are many important results that can be extracted from the COHERENT measurements. Using both timing and energy data, the flavor components of the flux can be measured.  Present estimates of the electron and muon flavor components are shown in Figure~\ref{fig:flavors}. It is expected that future data sets will be able to fully separate the flavor components. 

\par The COHERENT results have provided the first measurement of the neutron distribution in CsI~\cite{Cadeddu:2017etk,Cadeddu:2019eta,Coloma:2020nhf}. The RMS measurement of the neutron radius is $R_n = 5.0 \pm 0.7$ fm, assuming the Helm parametrization of the form factor. The measurement of the neutron skin is $R_n - R_p = 0.2 \pm 0.7$ fm, which is in agreement with the theoretical nuclear model predictions (see e.g. Fig.~4 in Ref.~\cite{Coloma:2020nhf}).

\bigskip
\noindent 
\underline{Lujan}: The Lujan Center's tungsten spallation target can be a prolific source of neutrinos from decays of stopped pions and muons created by an 800~MeV proton beam at Los Alamos National Laboratory. A 10 ton liquid argon scintillation detector or Coherent CAPTAIN-Mills (CCM) detector is built to study neutrino's coherent elastic scattering with argon nuclei. A ton-scale mass and a keV-range energy threshold may allow the CCM detector to possess leading sensitivity to potential low-mass dark-matter signals. 

\bigskip
\noindent 
\underline{ESS}: The European Spallation Source (ESS) will soon provide the most intense neutron beams for multi-disciplinary science. It will also generate the largest pulsed neutrino flux suitable for the detection of CE$\nu$NS~\cite{Baxter:2019mcx}.

\bigskip
\noindent 
\underline{JSNS$^2$}: The Japan Spallation Neutron Source of J-PARC is featured by an 1 MW beam of 3 GeV protons incident on a mercury target, creating an intense neutrino flux from the stopped-pion and stopped-muon decays.
The JSNS$^2$ (J-PARC Sterile Neutrino Search at J-PARC Spallation Neutron Source) experiment aims to search for the existence of neutrino oscillations and to offer the ultimate test of the LSND anomaly at a 17-ton fiducial volume Gd-dopped liquid scintillation detector~\cite{Ajimura:2017fld}. 
A new detector is being planned to study not only CE$\nu$NS but potential low-mass dark-matter signals~\cite{Ajimura:2020qni}. 

\bigskip
\noindent 
\subsection{Reactors}
The current theme of reactor experiments is on the observation of neutrino-nucleus elastic scattering at the kinematic regime where complete quantum-mechanical coherency is expected~\cite{Kerman:2016jqp}. The following experiments are either in data-taking mode, or are being planned, at different reactors sites around the world. 

\bigskip
\noindent 
\underline{CHILLAX}: The CoHerent Ionization Limit in Liquid Argon and Xenon (CHILLAX) project is an experimental effort to develop a xenon-doped argon ionization detector that can enjoy the benefits of both argon and xenon~\cite{CHILLAX_M7_2021}. Thanks to the relatively small atomic mass, an argon atom can pick up more kinetic energy from neutrino scatters than heavier elements can, and by doping it with xenon -- which has lower excitation/ionization energy and faster scintillation than argon does  -- the detector can be more efficient in producing ionization electrons from \cevns\ interactions and also generate detectable light signals with long wavelength and fast decays. Combining an argon target with a xenon detector-like performance, CHILLAX aims to develop the ideal large-mass noble liquid \cevns\ detector. With an expected energy threshold of 200-300eV, CHILLAX may detect a few times less \cevns\ interaction signal per kilogram than what is possible in eV-threshold detectors, but thanks to the scalability of the noble liquid technology CHILLAX can easily achieve an active mass of tens of kilogram and be a leading competitor in rate-oriented \cevns\  applications. 
CHILLAX is currently focusing on developing the first generation prototype detector. Once the xenon-doping benefits and low-energy sensitivity are experimentally demonstrated, we plan to build a $\sim$50kg detector to deploy either at the SNS (for BSM physics studies) or near a reactor (for sterile neutrino search and reactor monitoring demonstrations). 

\bigskip
\noindent 
\underline{CONNIE}: The Coherent Neutrino-Nucleus Interaction Experiment (CONNIE) uses low-noise fully depleted charge-coupled devices (CCDs) with the goal of measuring low-energy recoils from CE$\nu$NS of reactor antineutrinos with silicon nuclei~\cite{Aguilar-Arevalo:2019jlr}. The CCD detectors  can operate at a nuclear recoil threshold of approximately 30 eV, where the conversion from electron equivalent to silicon recoil energy is given by the so-called quenching
factor, from which measurements at low energies ($\approx 0.7$ keV) and new theoretical approaches are used \cite{QFmeasurementsSi, QFtheorynew, Sarkis2021}.

\par
\begin{figure}[!htb]
\centering
\includegraphics[width=2.1in,angle=270]{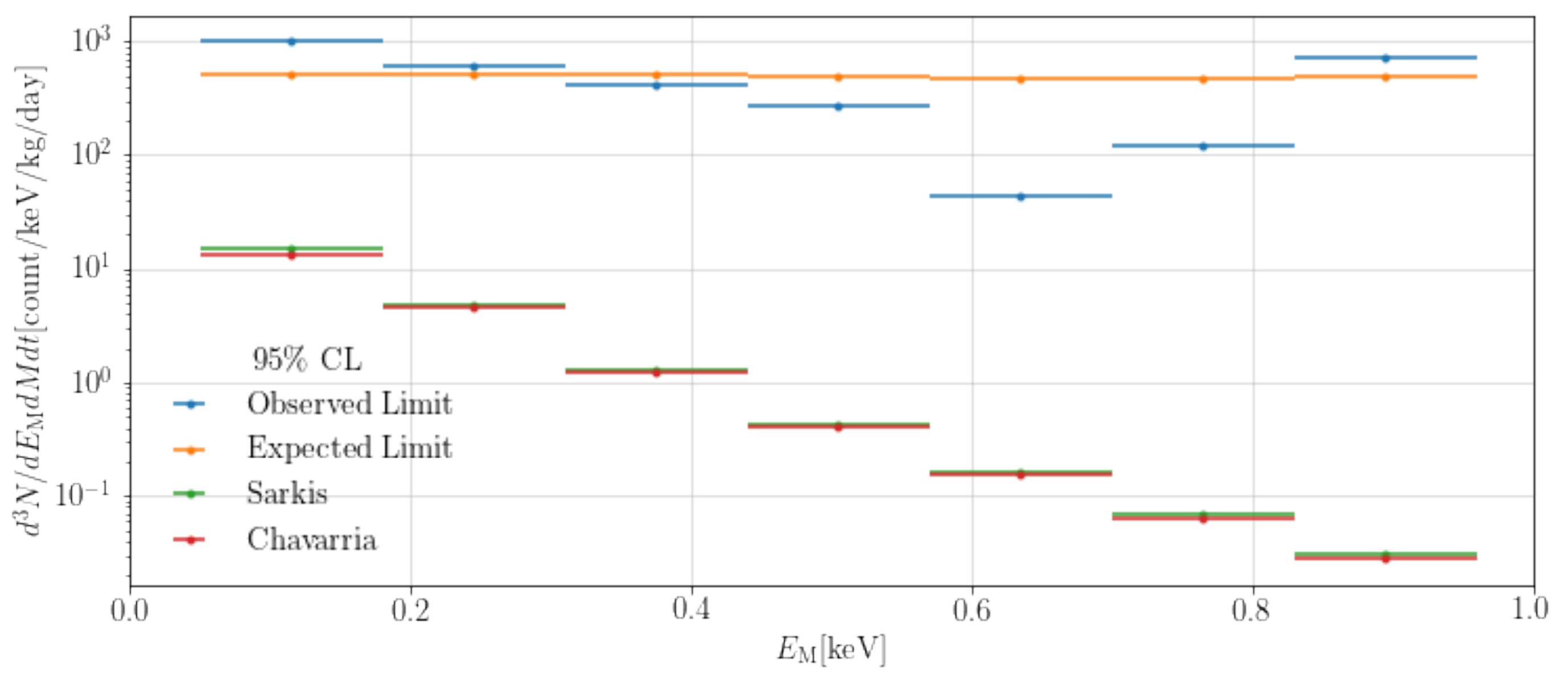}
\caption{CE$\nu$NS limit from CONNIE compared to the Standard Model predictions. Figure reproduced from Ref~\cite{CONNIE:2021ngo}.} 
\label{fig:connie}
\end{figure}

 CONNIE has reported results from its analysis with a detector array of 8 CCDs with a fiducial mass of 36.2 g, and a total exposure of 2.2 kg-days. In an analysis of the difference between the reactor-on and reactor-off spectra, no excess events are found at low energies, yielding upper limits at 95\% confidence level on the~\cevns~rate. In the lowest-energy range analyzed by CONNIE, $50-180$ eV, the expected limit is 34 (39) times the standard model prediction, depending on whether the Sarkis or the Chavarria quenching factor is assumed. The CE$\nu$NS limits from CONNIE are shown in Figure~\ref{fig:connie}. 

\par In addition, the upper bound from the previous analysis~\cite{Aguilar-Arevalo:2019jlr} in the lowest-energy bin was used to impose competitive constraints on the parameter space of some NSI models. CONNIE used two simplified extensions of the Standard Model with light mediators and obtained new world-leading constraints~\cite{CONNIE:2019xid} for vector mediator masses $M_{Z'}<10$~MeV and scalar mediator masses $M_{\phi}<30$~MeV. These results were quoted as a community milestone, being the first competitive BSM constraint from CE$\nu$NS at reactors.

\par In the next planned upgrade, CONNIE aims to lower the detection threshold, decrease the readout noise and increase the efficiency by upgrading the experiment with the recently demonstrated skipper-CCD sensor. The experiment is currently commissioning running with two skipper CCDs and optimising the data acquisition and analysis strategy. Studies show that skipper CCDs allow to lower the detection threshold to about 10 eV, increasing the neutrino rate about 6 times at the current experiment position. With the preliminary background rate measurement of 4~kdru, if the experiment is upgraded to 100~g mass, it could detect \cevns in about three months. The possibilities to reduce the background rate and increase the neutrino flux are being studied if the detector is moved inside the reactor dome, which would greatly improve its sensitivity.

\bigskip
\noindent 
\underline{CONUS}: The CONUS (COherent elastic Neutrino nUcleus Scattering) experiment employs four 1\,kg low energy threshold high-purity point-contact Germanium detectors to look for CE$\nu$NS. The experiment is located at the commercial nuclear power plant of Brokdorf, Germany, in a distance of 17.1\,m from the reactor core with a maximum thermal power of 3.9\,GW. With the electrically cooled spectrometers an energy resolution of 150-160eV (ionization energy, full width at half maximum ionization energy) at the 10.4\,keV K-shell X-ray of $^{71}$Ge/$^{68}$Ge is achieved \cite{bonet2021large}. The energy threshold is $\leq$1.875\,keV for nuclear recoils. With the onion-like shield consisting of layers of lead, borated polyethylene and a muon anti-coincidence veto a background level of $\sim$10\,counts/kg/d below 1\,keV ionization energy is achieved at reactor site \cite{bonet2021full}.
With the data collected in 2018 and 2019 an upper limit on CE$\nu$NS was derived in dependence of the quenching factor (ratio between detected ionization energy and recoil energy) in germanium \cite{bonet2021constraints}. For a quenching factor of k=0.16, this corresponds to a limit of 0.34\,kgd$^{-1}$ at 90\% confidence level (factor 17 above the standard model prediction). From the same data, limits on various BSM models \cite{CONUS:2021dwh} as well as the neutrino magnetic moment (from neutrino electron scattering) \cite{bonet2022first} were evaluated. With the data collected afterwards significant improvements on these limits are expected.
The quenching factor for germanium at low recoil energies is not well known and large discrepancies between the existing measurements persist. Recently, the CONUS collaboration carried out an own quenching measurement \cite{bonhomme2022direct} to significantly reduce this major systematic uncertainty. 
The final data collection run started in 2021 with several upgrades enhancing the sensitivity for a CE$\nu$NS detection. Significant improvements on the stability of the environment parameters were achieved. For the first time pulse shape information are collected as well with the intent to discriminate signals from background. The nuclear power plant shut down permanently at the beginning of 2022, which provides the opportunity to reduce the statistical uncertainty and improve the background understanding.

\bigskip
\noindent 
\underline{Dresden}: The experiment consists of a low-noise 3\,kg p-type point contact germanium detector which has been installed and operated at the Dresden-II power reactor, near Chicago, at about 10 meters from its 2.96 GW$_{th}$ core~\cite{Colaresi:2021kus}. The detector are enclosed by a compact shielding made of several layers of active vetos and passive shielding material. The results from an upgraded setup are interpreted to have a preference for a \cevns component in the data, when being compared to a background only model~\cite{Colaresi:2022obx}.

\bigskip
\noindent 
\underline{MINER}: The Mitchell Institute Neutrino Experiment at Reactor (MI$\nu$ER) experiment was launched to use cryogenic germanium and silicon detectors with a low nuclear recoil energy thresholds to register nuclear recoils of coherent elastic neutrino-nucleus scattering (CE$\nu$NS) at a TRIGA research nuclear reactor at the Texas A\&M University~\cite{Agnolet:2016zir}. This reactor has a movable core (1\,m to 10\,m) that will allow precision studies of very short baseline neutrino oscillation by comparing rates as a function of distance and largely eliminating reactor flux uncertainties. Close proximity of the detector to the reactor core, combined with multiple low threshold detectors with event-by-event discrimination between the dominant electromagnetic background and the nuclear recoil signal provides sensitivity to BSM physics, sterile neutrinos that oscillate away on a few-meter scale, and above all a highly sensitive probe for applied reactor monitoring for safeguards and non-proliferation. Planned deployment of the MIN$\nu$ER experimental set up at the South Texas Project (3 GW) power reactor will provide significant further improvement in measurement sensitivity. 

\bigskip
\noindent 
\underline{NEON}: The Neutrino Elastic-scattering Observation on NaI(Tl) (NEON) experiment uses high-light yield NaI(Tl) crystals to observe the CE$\nu$NS events at a distance of 24\,m from the core of the Hanbit nuclear reactor in Korea. 
Commercial reactors produce a large number of anti-neutrinos with a thermal power of 2.8\,GW, and the on-going NEOS sterile neutrino program~\cite{Ko:2016owz} provides a systematic understanding of the environment. 
This experiment utilized the previous experiences of the NaI(Tl) crystal detectors for the COSINE-100 dark matter search experiments~\cite{Adhikari:2018ljm,Adhikari:2019off}.  
The NaI(Tl) detector used in the COSINE-100 experiments showed a light yield of 15-photoelectrons/keV~\cite{Adhikari:2017esn}, and a multivariate machine larning technique was used to effectively remove the noise event caused by PMT to reach a low energy threshold of 1~keV~\cite{COSINE-100:2020wrv}. Preliminary studies to lower the energy threshold achieved 0.5~keV energy event access with 80\% selection efficiency and 25\% noise contamination level.  This can be improved by adopting deep machine learning algorithm currently in development. 
In addition, we developed a novel encapsulation method of the NaI(Tl) crystals with improved light collection efficiency up to 22 photoelectrons/keV~\cite{Choi:2020qcj}. Energy thresholds of less than 0.3\,keV can be achieved with these detectors. 
The first phase NEON experiment~(NEON-pilot) was built with a $2\times3$ array of 6 detectors with a total mass of 15~kg using the commercial quality crystals while the next phase experiment~(NEON-1) may use up to 100\,kg of the low-background NaI(Tl) crystals~\cite{Park:2020fsq}. 
The NEON-pilot crystals were immersed in an 800-L liquid scintillator. It was shielded with 10-cm-thick leads and 30-cm-thick polyethylene. The shields and DAQ systems closely follows the COSINE-100 dark matter experiment~\cite{Adhikari:2017esn,Adhikari:2018fpo}. The NEON-pilot experiment is currently being commissioned on the ground site and is expected to be installed at the reactor site by the end of 2020. We plan to take 1 year of reactor-on data with 5 months of off data expecting the observability of the CE$\nu$NS process with a significance of more than 3$\sigma$ and will upgrade it to NEON-1 for more than 5$\sigma$ observation. 

\bigskip
\noindent 
\underline{NUCLEUS}: The NUCLEUS experiment aims for a detection of CE$\nu$NS using CaWO$_4$ cryogenic calorimeters with nuclear recoil thresholds around 20~eV. This unique feature demonstrated in an early prototype~\cite{Strauss:2017cuu,Strauss:2017cam,Angloher:2017sxg} will allow observation of the majority of tungsten recoils induced by reactor antineutrinos by accessing unprecedentedly low energies, taking full advantage of the coherent cross-section boost. The first experimental phase will deploy a 10~g cryogenic target~\cite{Rothe:2019aii} composed of approximately 6~g CaWO$_4$ and 4~g Al$_2$O$_3$ in a new experimental location~\cite{Angloher:2019flc} at the Chooz nuclear power plant in France. The two target materials feature widely different CE$\nu$NS cross-sections but a comparable neutron response, useful for in-situ measurement of potentially dangerous nuclear-recoil backgrounds.\\ 
The experimental setup will consist of a dry dilution refrigerator, a compact passive shielding made of neutron moderators and lead, active muon~\cite{Wagner:2022iqf} and gamma anticoincidence veto detectors as well as an integrated LED-based calibration system. The setup is under construction for commissioning in Munich in 2022 before deployment at the reactor site planned from 2023 on. 

\bigskip
\noindent 
\underline{nuGen}: The nuGEN experiment aims at the detection of \cevns with low-background, low-threshold HPGe detectors installed at a distance of $\sim10$\,m to one of the 3.1\,GW$_{th}$ reactors at the Kalinin Nuclear Power Plant (KNPP) in Russia. The detectors are surrounded by a compact passive shielding and an active muon veto, which are installed on a movable platform to modify the distance to the reactor core. nuGEN has been installed on-site in 2019 and preliminary data from first science runs have been presented~\cite{nuGEN_mag7}. 

\bigskip
\noindent 
\underline{NUXE}: The NUXE experiment will use a liquid xenon detector to observe reactor neutrino CE$\nu$NS events down to single ionization electron signals~\cite{Ni:2021mwa}. The experiment is currently under development at UC San Diego with a 30-kg liquid xenon target in an electron counting chamber (ECC). Major effort is reducing the background down to the single electrons, corresponding to a nuclear recoil energy threshold of $\sim$300~eV~\cite{Lenardo:2019fcn}. 

\bigskip
\noindent
\underline{PALEOCCENE}: The PALEOCCENE concept~\cite{Cogswell:2021qlq,Alfonso:2022meh} aims to exploit the crystal defects caused by nuclear recoil by using an optical readout scheme based on the imaging of individual color centers. The resulting detectors would be room-temperature, passive devices with recoil thresholds close to the threshold damage energy of the detector material of 100\,eV or less. The project is in the early stages of R\&D.

\bigskip
\noindent 
\underline{Ricochet}: The Ricochet neutrino experiment aims to measure neutrinos produced from nuclear reactors by using cryogenic bolometers to identify the signature nuclear recoil from CE$\nu$NS\xspace~\cite{Ricochet:2021rjo}.  In order to overcome the high level of electromagnetic background present at low energies, Ricochet will make use of particle identification in order to discriminate between electron and nuclear recoils.  

The future Ricochet experiment will be deployed at the ILL-H7 site in Grenoble, France. The H7 site starts at about \SI{8}{\m} from the ILL reactor core that provides a nominal nuclear power of \SI{58.3}{\MW}, leading to a neutrino flux at the Ricochet{} detectors \SI{8.8}{\m} from the reactor core of about 1.2$\times$10$^{12}$~cm$^{-2}$s$^{-1}$. The reactor is operated in cycles of typically 50 days duration with reactor-off periods sufficiently long to measure reactor-independent backgrounds, such as internal radioactivity or cosmogenic induced backgrounds with high statistics.  It is located below a water channel providing about \SI{15}{\m} water equivalent (m.w.e.) against cosmic radiation. It is not fed by a neutron beam and is well-shielded against irradiation from the reactor and neighboring instruments (IN20 and D19). The site is well-characterized in terms of backgrounds, and the operation of the STEREO neutrino experiment at this site has been successfully demonstrated~\cite{STEREO:2018blj}.

The experiment will make use of two detector technologies/targets.
The CryoCube will consist of an array of 27 ($3\times3\times3$) high purity germanium crystal detectors, encapsulated in a radio-pure infrared-tight copper box suspended below a lead shield inside the crysotat~\cite{RICOCHET:2021gkf}. Each detector mass is about 38\,g to reach a total target mass around one kilogram. A low-energy threshold is desired as the discovery potential scales exponentially with lowering the energy threshold. Considering a 50\,eV energy threshold, about 12.8\,evts/kg/day of CE$\nu$NS\xspace interactions is expected in the CryoCube detector array.
To reach such threshold, the CryoCube detectors will be equipped with germanium neutron transmutation doped sensors (NTD). To achieve particle identification, the detectors will have a double heat and ionization readout. Ionization measurement is realized thanks to aluminum electrodes allowing to apply an electric field and collect signals from the ionization electron-hole pairs drifting across the crystal. With an anticipated particle identification threshold of about 100\,eV, thanks to the combination of a 10\,eV and 20\,eVee (electron-equivalent) heat and ionization baseline resolutions (RMS), the CryoCube detector array should lead to a CE$\nu$NS\xspace detection significance after one ILL reactor cycle (50-days) between 4.3--17.3~$\sigma$, depending on the final background level achieved.

Q-Array -- the complimentary detector array within Ricochet -- will consist of 9 cubes of superconducting zinc cubes, each with a mass of about 35\,g, as its target. Using superconductors as the primary detector is a novel technology which is expected to provide detection sensitivity theoretically down to the Cooper pair binding energy.  The expected discrimination mechanism begins with the different efficiency of quasiparticle (QP) production (breaking Cooper pairs) by electron recoils (higher QP production) vs.~nuclear recoils (lower QP production).  The initial phonon production from the recoil is followed by a slower phonon production as QPs relax to the ground state.  The relative ratio between initial phonons and QP-induced phonons thereby gives rise to a usefully discriminating pulse shape.  Transition edge sensors (TES) will be used for the readout of the phonon signals from these superconducting bolometers. Initial prototype TES chips with a transition temperature of \SI{80}{\milli\K} were developed by Argonne National Laboratory for this use. Using results from these sensors, a second version was designed and fabricated, with a transition temperature of $\sim$20~mK. These devices are under testing~\cite{RanLTD} and a significant decrease in sensor threshold is expected. 

The background level at the ILL site is expected to be high, due to the proximity to the nuclear reactor core ($\approx$8~meters), the neighboring experiments emitting large amounts of gammas and neutrons (IN20 and D19). Despite the presence of the $\approx$15 m.w.e. artificial overburden provided by the water transfer channel of the reactor directly above the experiment, the site remains exposed to cosmic radiation. As we expect to observe about 12.8 CE$\nu$NS\xspace events/kg/day, a highly efficient background mitigation strategy is mandatory. As a result, Ricochet aims for an electronic recoil background at the level of 100~events/day/keV/kg.  Such an electronic background level should be efficiently rejected thanks to our detectors' particle identification capabilities. However, as such discrimination doesn't hold for neutron induced nuclear recoil, the latter are expected to be our ultimate background. We are therefore aiming for a nuclear recoil background level around 5~events/kg/day to ensure a favorable CE$\nu$NS\xspace signal to noise ratio.  

Ricochet is currently scheduled to see ``first light" in 2023.

\bigskip
\noindent 
\underline{RED-100}: RED-100 is a two-phase xenon emission detector built to observe the coherent elastic scattering of reactor electron antineutrinos off xenon atomic nuclei~\cite{Akimov:2017hee}. The mass of the detector medium is 160 kg in the sensitive volume, and about 100 kg in the fiducial volume --- the largest value among detectors developed for CE$\nu$NS observations at reactors. The capability of detector full-scale operation in the background conditions caused by cosmic radiation has been demonstrated in a ground-level laboratory~\cite{Akimov:2019ogx}. The detector was deployed at the Kalinin NPP in 2021 at 19~m distance under the 3~GW reactor core (50~m.w.e. overburden) and acquired both reactor on and off data in January and February of 2022 with an energy threshold of about 1~keV$_{nr}$.

\bigskip
\noindent 
\underline{SBC}: The SBC (Scintillating Bubble Chamber) Collaboration is developing novel liquid-argon bubble chambers for GeV-scale dark matter and CE$\nu$NS physics~\cite{Giampa:2021wte}. The first detector, SBC-Fermilab, with 10-kg active mass, is currently under construction for characterization and calibration in the NuMI tunnel at Fermilab, aiming to reach a threshold for nuclear recoils of 100~eV. The detector consists of a quartz jar filled with superheated liquid argon, which is spiked with ppm levels of xenon acting as a wavelength shifter. Cameras are used to image bubbles, silicon photo-multipliers detect scintillation, and piezo-acoustic sensors listen for bubble formation. A duplicate detector constructed with low-background components, SBC-SNOLAB, will follow for a search for 0.7--7~GeV dark matter.

Following this initial program, a deployment of one of these detectors after $\sim$2024 at a nuclear reactor could make a high signal-to-background measurement of reactor neutrino CE$\nu$NS. Sensitivity to the weak mixing angle, neutrino magnetic moment, and a Z’ gauge boson mediator have been calculated for deployments at both a 1~MW research reactor and a 2~GW power reactor~\cite{SBC:2021yal}. Background characterizations are ongoing at the 1~MW TRIGA Mark III research reactor located at the National Institute for Nuclear Research (ININ) near Mexico City for a potential first reactor deployment 3~m from the core.

\bigskip
\noindent 
\underline{TEXONO}: 
The TEXONO collaboration has been studying neutrino physics with sub-keV
germanium detectors at the Kuo-Sheng Reactor Neutrino Laboratory (KSNL) in Taiwan~\cite{Wong:2016lmb}. There is a national policy of de-commissioning nuclear power in Taiwan, and the
Kuo-Sheng Reactor will be phased out by 2023. As a result, there are no plans on
expansion or new projects to the KSNL program. The collaboration would seek to
continue the studies via collaboration with other existing reactor laboratories.

Data taking and R and D program are conducted with electro-cooled (EC) point-contact germanium detectors (PCGe). As of summer 2020, detector mass up to 1.43 kg
are built and threshold as low as 200 eV ee is achieved. The data would also bring
improved sensitivities to the searches on various Beyond Standard Model (BSM)
physics channels, such as neutrino magnetic moments~\cite{Wong:2006nx} and milli-charged neutrinos~\cite{Chen:2014dsa}. Active theory program is being pursued in parallel, with focuses on atomic
corrections to $\nu$N (and $\chi$N for dark matter) cross-sections~\cite{Chen:2014dsa}, as well as BSM searches.

\bigskip
\noindent 
\underline{NEWSG}: 
NEWS-G is a direct dark matter detection experiment, sensitive to  light Dark Matter (DM) between 0.1 and \SI{10}{GeV/c^2}. NEWS-G uses a spherical gaseous detector, the Spherical Proportional Counter (SPC)~\cite{Giomataris:2008ap}. The detector is composed of a spherical shell made of radio-pure copper, acting as the cathode, and a read-out sensor at the center with either a single anode~\cite{Katsioulas:2018pyh} or multiple anodes~\cite{Giomataris:2020rna, Giganon:2017isb}. The detector is versatile and can operate with a wide variety of gas mixtures. Use of light elements, such as H, He, C, and Ne, is ideal for light DM searches thanks to the kinematic match between target and projectile, and the favorable ionization quenching factor compared to heavier elements. The SPC exhibits several features, such as:
  simple, few component build;
   very low energy threshold (single ionization electron level) independent of detector size~\cite{Bougamont:2010mj}; and
  fiducialization and event discrimination through pulse shape analysis.
These make it an invaluable tool for searches requiring detection of low-energy recoils. The first NEWS-G results were published in 2017~\cite{Arnaud:2017bjh} with a 60\,cm diameter SPC installed at LSM~\cite{Arnaud:2017bjh}, producing the most stringent limits for DM masses below \SI{600} {MeV/c^2} at that time. Currently (2020), the advanced 140\,cm diameter SPC~\cite{Katsioulas:2020ftw,Giroux:2019lvw} is being installed at SNOLAB following a commissioning phase at LSM in 2019. 

The SPC's features that make it ideal for light DM searches make it also appealing for the detection of neutrinos through CE$\nu$NS. The study of CE$\nu$NS with a sub-keV energy threshold detector, like the SPC, allows for a rich physics program and opens a window to physics beyond the Standard Model that will appear as deviations from the expected recoil spectra predicted by SM interactions~\cite{Scholberg:2005qs}. For example, the measurement of a non-zero neutrino magnetic moment \cite{Scholberg,Kosmas:2015sqa} and the search for sterile neutrinos. The use of SPC for CE$\nu$NS detection can lead to several applications related to nuclear reactors, such as monitoring reactor neutrino fluxes, probing the reactor fuel to control plutonium production and the study of reactor anti-neutrino energy spectrum, which is not well understood below the Q-value of the inverse beta decay process. Practically, taking into account the detector energy threshold (\SI{50}{eV_{ee}}) and the SPC size (\SI{60}{cm}), Ar and Ne are the strongest candidates to be used as targets, with event rates of 15 and \SI{11}{events/kg/day/GW} respectively, however, this does not exclude the use of other gas mixtures for specification applications. 

NEWS-G plans to construct a \SI{60}{cm} diameter sphere made of ultra-pure copper for CE$\nu$NS studies. The detector will be encased in a shielding inspired from the GIOVE~\cite{giove} and CONUS~\cite{conus} experiments. A muon veto will complete the shielding apparatus. The experimental setup will be installed at Queen's University, Kingston, Canada, to assess the environmental and cosmogenic backgrounds and to establish if any alterations are required for operations in the proximity of a reactor. The construction of the shield will start in fall 2020 and commissioning at Queen's University is expected to take place by summer 2021. The background studies will be conducted through 2021 and 2022, while a reactor site is identified. One year of data taking is planned after the installation of the detector at a reactor site.

\subsection{Dark matter \& CE$\nu$NS detectors}

As described above, neutrinos from the astrophysical sources like the Sun and supernovae as well as from the atmospheric cosmic-ray-showers produce nuclear recoils via CE$\nu$NS, leading to signals in future dark matter detectors. 
The eventual presence of an unshieldable background in multiton-scale dark matter detectors has been anticipated for some time~\cite{Cabrera:1984rr,Drukier:1986tm,Monroe:2007xp,Vergados:2008jp,Strigari:2009bq,Gutlein:2010tq,Gelmini:2018ogy}, and is thought to present a major obstacle for improving the sensitivity of these experiments. The central problem is that for many of the most commonly sought-after dark matter models produce nuclear recoil signals that look remarkably similar to the CE$\nu$NS recoil energy spectra generated by natural neutrino sources. Due to the finite systematic uncertainty on the fluxes of those sources eventually a feeble-enough DM signal could disappear under the expected variation in the neutrino event rate, and because its signal would not be distinct enough from the background no positive identification of dark matter would be possible. Naively this implies that there is a ``floor'' to the sensitivity of direct detection experiments~\cite{Billard:2013qya}. 

\begin{figure}[!htb]
\centering
\includegraphics[height=1.8in]{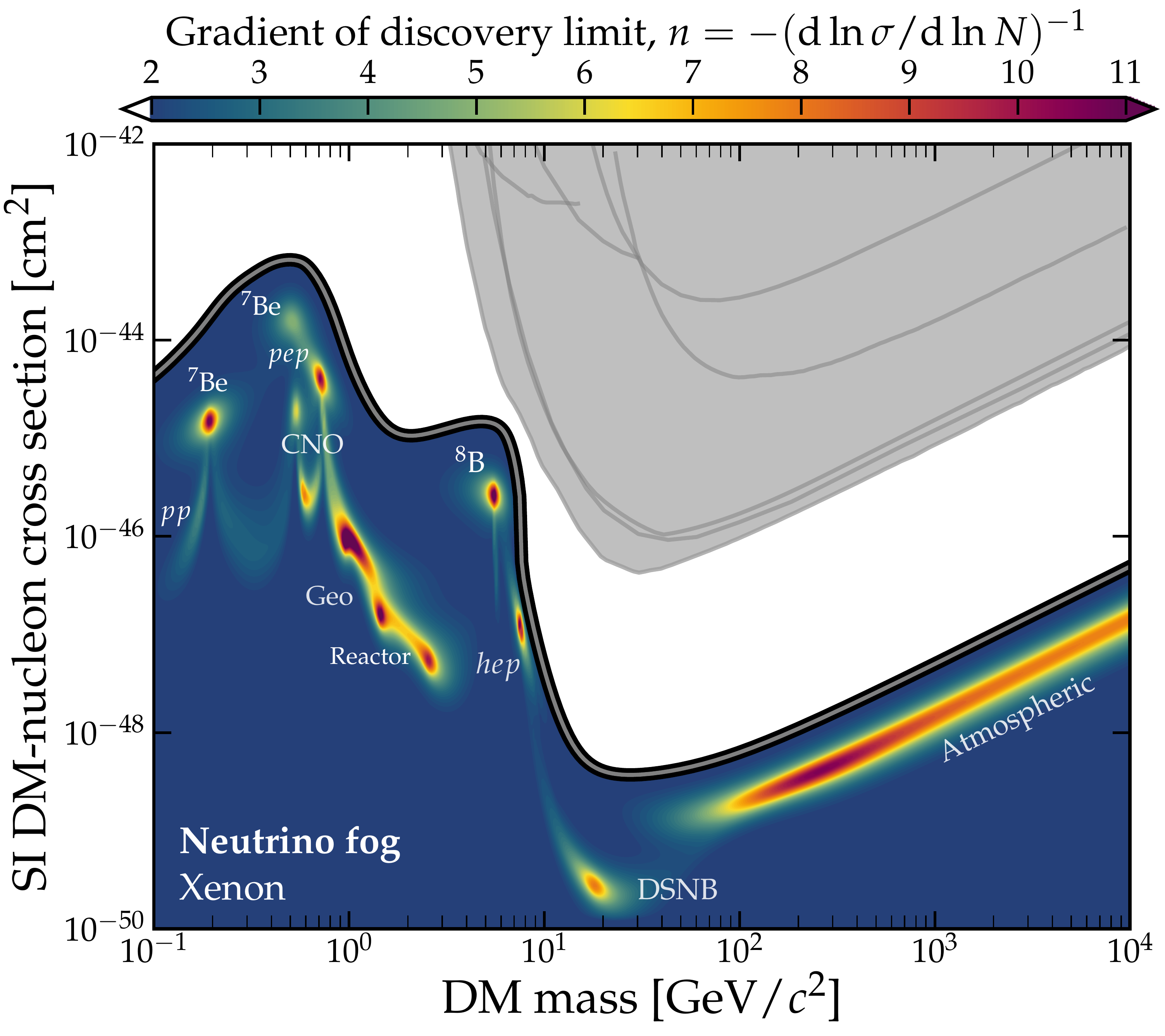}
\includegraphics[height=1.5in]{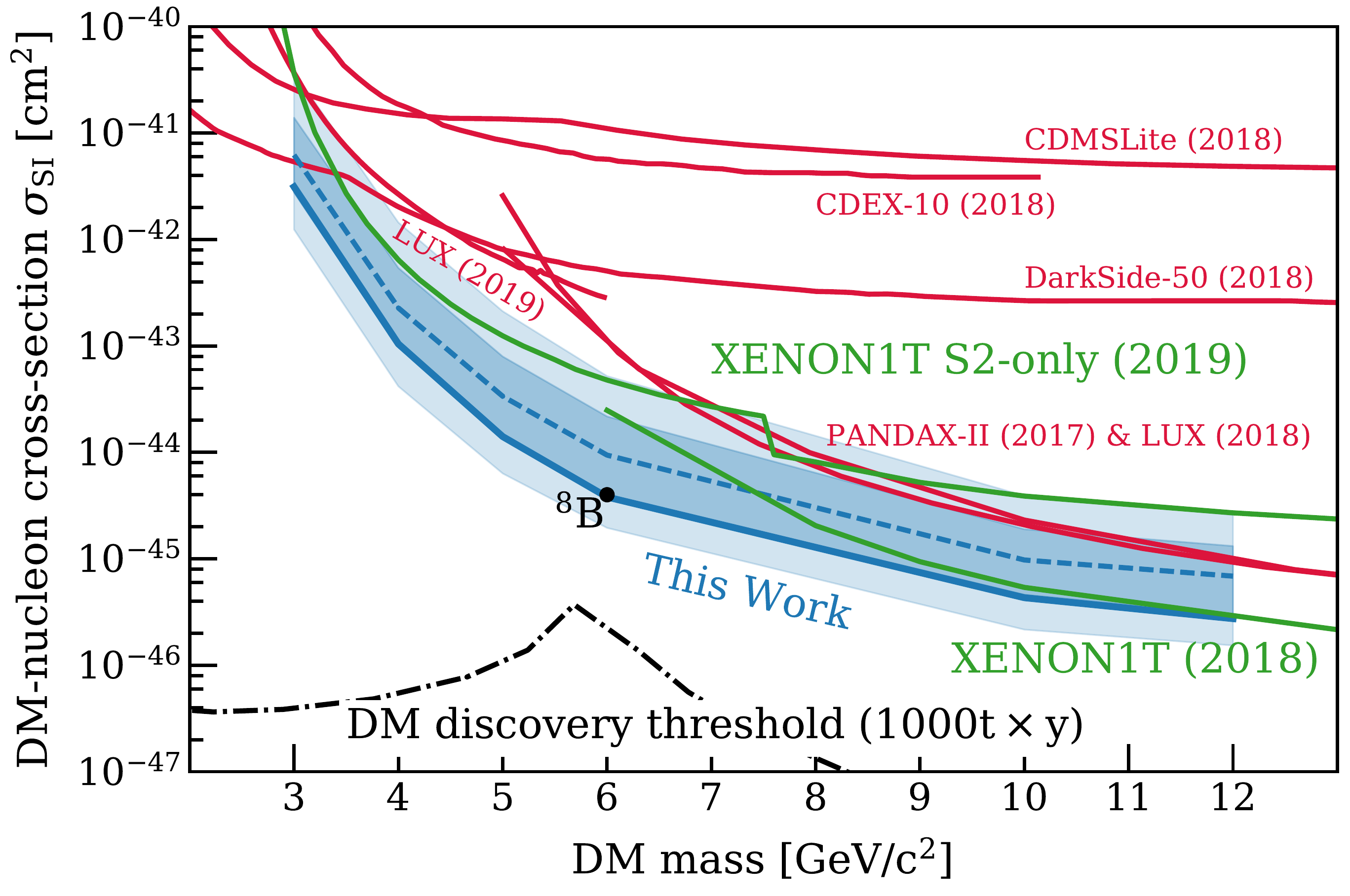}
\caption{Left: The neutrino ``fog'' for a xenon direct DM experiment (adapted from Ref.~\cite{OHare:2021utq}). The opacity of the neutrino fog, i.e.~how much the CE$\nu$NS signal impedes DM discovery, is encoded by the value of $n$ shown by the colorbar. This is the exponent with which a DM discovery limit scales with exposure i.e. $\sigma \propto (MT)^{-1/n}$, so $n<2$ in an approximately background-free regime, $n = 2$ under standard Poissonian background subtraction, and $n>2$ when there is saturation of the signal by the background. The level of severity of this background saturation varies with the DM mass and cross section, the fluxes of each neutrino source, and the uncertainty on those fluxes. The relevant source for each part of the parameter space is labelled. Right: $^{8}$B solar neutrino flux limit from XENON1T, for an exposure of 0.6\,tonne$\times$year~\citep{Aprile:2020thb}.} 
\label{fig:8BXenon}
\end{figure}

Since the dark matter and CE$\nu$NS signals are not exactly identical--- it is not impossible to search for dark matter in the presence of a CE$\nu$NS background, but it does entail significant reduction in experiment's sensitivity. This observation has been called ``neutrino fog''~\cite{OHare:2021utq}. One approach for visualising the nature of this fog is shown in the left-hand panel of Fig.~\ref{fig:8BXenon}, where the color displayed encodes the ``opacity'' of the neutrino fog in different parts of the spin-independent DM-nucleon cross section parameter space. One can appreciate here that major sources of neutrino that need to be understood better to aid the search for dark matter are $^7$Be, $^8$B and atmospheric neutrinos, as the height of this fog in cross section is crucially dependent on the size of the systematic uncertainty on those fluxes~\cite{Ruppin:2014bra}. Neutrino floor considerations play an important role when considering the feasibility of future experiments, as decreased sensitivity to DM signals in vicinity of neutrino background also hinder the increase in experimental sensitivity with scaling in detector exposure~(e.g.~\cite{Dent:2016wor,Gelmini:2018ogy}). There are several approaches for circumventing the neutrino floor if an experiment can access additional information to further discriminate the dark matter signal and neutrino background, including the use of annual modulation~\cite{Davis:2014ama}, target complementarity~\cite{Ruppin:2014bra,Gelmini:2018ogy,Gaspert:2021gyj,OHare:2020lva}, and directionality~\cite{OHare:2015utx,Grothaus:2014hja,Mayet:2016zxu,OHare:2017rag,Franarin:2016ppr,Vahsen:2020pzb,Vahsen:2021gnb,Sassi:2021umf}. It should also be emphasized that any additional uncertainties in the neutrino signals~\cite{OHare:2016pjy,AristizabalSierra:2017joc,Gonzalez-Garcia:2018dep,Papoulias:2018uzy,AristizabalSierra:2021kht} beyond the systematic uncertainties on the flux will raise the height of the neutrino fog. Since the signals of different types of dark matter-nucleus interactions do not identically align with the CE$\nu$NS signal, depending on the scenario the neutrino fog could be much lower, or absent entirely~\cite{Dent:2016iht, Dent:2016wor,Gelmini:2018ogy,Essig:2018tss,Wyenberg:2018eyv}. For example, inelastic dark matter scattering that could be naturally associated with models of multi-component dark sectors could help lift the signal degeneracy of the neutrino floor~\cite{Gelmini:2018ogy}.

Many upcoming experiments will be sensitive to CE$\nu$NS and are expected to enter the neutrino fog, such as XENONnT~\citep{aprile2020projected}, LZ~\citep{akerib2020projected} and DARWIN~\citep{Aalbers:2016jon,Aalbers:2022dzr}. Here we discuss the several experiments that will be sensitive to~\cevns~in the forthcoming years. 

\bigskip
\noindent 
\underline{XENON}: 
The XENON1T experiment operated a dual-phase time projection chamber (TPC) filled with 3.2 tonnes of ultra-pure liquid xenon (LXe). The TPC contained 2.0 tonnes of LXe that is sensitive to ionization electrons (S2) and scintillation photons (S1) produced by interactions therein. In a fiducial volume with 1.0 tonnes of LXe, a background level down to $(76\pm2)\,\mathrm{events}/(\mathrm{tonne}\times\mathrm{year}\times\mathrm{keV})$ has been achieved. A WIMP dark matter search lasted from December 2017 to February 2018. Using this data, the XENON collaboration has performed a first sensitive search for solar $^{8}$B neutrinos through nuclear recoils from the CE$\nu$NS process~\citep{Aprile:2020thb}. $^8$B CE$\nu$NS leads to an average nuclear recoil energy of $\sim$\,1\,keV, requiring unprecedented low energy thresholds in identifying scintillation and ionization signals. In this analysis, the threshold in ionization and scintillation signals was lowered down to 4\,electrons and 2\,photon-electrons, respectively. With an exposure of 0.6\,tonne$\times$year, the expected CE$\nu$NS signal is 2.1 events with a background expectation of 5.4 events, dominated by the accidental pileup of isolated-S1 and isolated-S2 signals. The mean discovery power of $^8$B CE$\nu$NS is $2\,\sigma$, limited by the exposure of this experiment. No significant excess from $^8$B CE$\nu$NS is found in XENON1T. The data is used to constraint new physics beyond the standard model, such as WIMPs and non-standard neutrino interactions, as well as the response of LXe to low energy nuclear recoils. The new world leading WIMPs sensitivity is shown in Figure~\ref{fig:8BXenon}. 

The XENONnT experiment is currently operating at the Gran Sasso underground laboratory in Italy, aiming to acquire a 20 t~$\times$~y exposure~\citep{aprile2020projected}. As the isolated-S1 rate scales up with the
larger number of PMTs and the isolated-S2 rate with the detector surface area, the accidental coincidence (AC) background will be the biggest challenge for the discovery of $^8$B CE$\nu$NS. The AC background modeling and discrimination techniques used in the XENON1T analysis will be further developed to improve the sensitivity of XENONnT to $^8$B CE$\nu$NS. The novel cryogenic liquid circulation system developed to ensure efficient purification in XENONnT will mitigate the reduction of S2s due to impurities, improving the acceptance of low-energy NRs from $^8$B neutrinos. Additionally, the data will be analyzed in a trigger-less mode to minimize efficiency loss and to better understand the AC background. Together with the significantly larger exposure, these techniques give XENONnT a strong potential to discover $^8$B CE$\nu$NS.

\bigskip
\noindent 
\underline{LUX-ZEPLIN}: 
LUX-ZEPLIN (LZ) is a dual-phase xenon TPC with a 7-tonne active mass located 1 mile underground at the Sanford Underground Research Facility (SURF) in Lead, South Dakota. 
The LZ detector was designed to search for interactions of particle dark matter in the mass range from 1 GeV/c$^2$ to 10 TeV/c$^2$.
Because the \cevns\ process can produce low-energy nuclear recoil signals  similar to those produced by low-mass WIMPs, LZ will also be sensitive to astrophysical sources of neutrinos such as solar $^8$B neutrinos and atmospheric neutrinos. Over the full 15.34 tonne-year exposure, LZ is expected to observe 0.65 events from atmospheric neutrinos and 36 events from $^8$B neutrino 
in its WIMP search campaigns~\citep{akerib2020projected}. A positive detection of $^8$B \cevns signals will be an unambiguous confirmation of LZ's low-mass WIMP sensitivity. As the \cevns mechanism is insensitive to neutrino flavors, LZ's measured flux provides a  data point complementary to large solar neutrino experiments that rely on charge-current interactions. 

The observable rate of $^8$B \cevns in LZ  depends strongly on the detector energy threshold. LZ relies on the detection of scintillation and ionization signals produced by particle interactions in the active liquid volume. Scintillation in the liquid produces a prompt signal (S1), and the ionization electrons produce a delayed electroluminescence response (S2) in a thin gas region above the liquid. For nuclear recoils with the energy of 1 keV, only $\sim 3$ photons and $\sim 6$ electrons are produced on average with the nominal LZ drift field of $\sim 200$ V/cm\citep{Lenardo:2019fcn}. With a photon detection efficiency of $\sim 0.1$, the observable $^8$B \cevns event rate can increase by a factor of $\sim 5$ when the S1 signal requirement is relaxed from 3 PMT coincidence to 2. However, as we relax the detector thresholds, the rate of instrumental backgrounds will also rise as a result of phony S1s (ex. PMT dark count coincidence) and spurious S2s (ex. grid electron emission). Such backgrounds could obscure us from observing $^8$B neutrino clearly. Experimental techniques that can lower the threshold while mitigating background are being actively explored. 

In addition, LZ is capable of detecting supernova neutrinos within our Milky Way, which occurs at a rate of 1-3 per century. The burst of SN neutrinos, typically with kinetic energy of O(10 MeV), can produce a stream of \cevns interactions in liquid xenon within seconds. LZ's DAQ system was designed to keep a high up-time to measure the total energy released, and can generate a real-time trigger as a part of Supernova Neutrino Early Warning System (SNEWS) to the astrophysics community.

\bigskip
\noindent 
\underline{DARWIN}: 
DARWIN (DARk matter WImp search with liquid xenoN) is a proposed next-generation dark matter experiment that will operate 50\,t (40\,t active) of xenon in a cylindrical time projection chamber with 2.6\,m in diameter and height~\cite{Aalbers:2022dzr}. The TPC will be placed in a double-walled cryostat vessel surrounded by neutron and muon vetoes. While DARWIN's primary goal is to observe particle dark matter in the $\sim$1\,GeV-100\,TeV mass range, it will also be able to measure the solar $^{8}$B neutrino flux, as well as atmospheric and supernovae neutrinos via CE$\nu$NS. The expected $^{8}$B neutrino rate is $\sim$90 events/(t\,yr)~\cite{Baudis:2013qla,Aalbers:2016jon}, depending on the achieved energy threshold. The measurement of the atmospheric neutrino flux requires a large exposure of about 700\,t\,yr~\cite{Newstead:2020fie}. A DARWIN-like detector would be able to observe astrophysical neutrinos of all flavours from core-collapse~\cite{Lang:2016zhv}, as well as failed core-collapse and thermonuclear runaway fusion~\cite{Raj:2019sci} supernovae. Typically, about 100 events are expected  from a core-collapse supernova at a distance of 10\,kpc and a 27\,M$_{\odot}$ progenitor mass~\cite{Lang:2016zhv}. The detection of neutrinos from failed core-collapse supernovae would deliver the time when the proto-neutron star collapses into a black hole~\cite{Raj:2019sci} and, in conjunction with the detection of gravitational waves, would identify the progenitor of a failed supernova. Finally, a large xenon detector such as DARWIN would dramatically improve the sensitivity to the diffuse supernova neutrino background (DSNB) in the $\nu_x$ channel, where $\nu_x \subset (\nu_{\mu}, \nu_{\tau}. \bar{\nu}_{\mu}, \bar{\nu}_{\tau})$~\cite{Suliga:2021hek}. While there are strong upper limits on the $\bar{\nu}_{e}$ flux from Super-Kamiokande, of 2.7\,cm$^{-2}$s$^{-1}$, the limits on $\nu_x$ are about three orders of magnitude weaker. DARWIN would be able to reach a sensitivity of $\sim$10\,cm$^{-2}$s$^{-1}$ per flavour. While this is not sufficient for a detection, such a constraint would exclude many DSNB scenarios with new astrophysics or physics~\cite{Suliga:2021hek}.

\bigskip
\noindent 
\underline{SuperCDMS SNOLAB}: 
SuperCDMS SNOLAB is a dark matter search focused on the 0.5--5~GeV mass range~\cite{supercdms_sensitivity_2017, supercdms_snowmass_2022}.  It will use two kinds of cryogenic solid-state detectors. The first type, iZIP detectors, will have the capacity to discriminate nuclear recoils from electron recoils via measurement of athermal phonons and ionization production down to 1--2~keV recoil energy.  The second type, HV detectors, will use a high drift field to transduce the charge signal into athermal phonons, provide a recoil energy threshold about 10$\times$ lower, though without the ability to discriminate nuclear recoils.  These thresholds should enable the detection of $^8$B neutrino \cevns in a solid-state detector for the first time, albeit with low statistics.  The experiment is currently under construction at SNOLAB and anticipates beginning to acquire data in late 2023.



Going forward, the SuperCDMS Collaboration anticipates extending its scientific reach for dark matter with the SNOLAB facility primarily through detector improvements that will provide access to much lower energy recoils~\cite{supercdms_snowmass_2022}.  (Modest background upgrades will also be implemented.)  This work will provide sufficiently low threshold to detect \cevns of solar neutrinos from the CNO, $pep$, $^7$Be, and even $pp$ reaction chains, going well past the $^8$B neutrinos detectable in SuperCDMS SNOLAB.  In particular, with the 0.5~eV threshold anticipated for 25-gram Si detectors operated with phonon-only readout at 0V bias (i.e., neither iZIP nor HV, and smaller in mass than the kg-scale SuperCDMS SNOLAB detectors), the rate of solar neutrino events will be roughly 0.01/kg-day.  With anticipated exposures of 12--240 kg-yr, there will be the potential to detect tens to hundreds of $pp$ chain \cevns events.  These events will compete with a background of coherent photonuclear scattering, which can be modeled well based on measurements of Compton scattering at high energies, and with environmental backgrounds such as vibrations, RF noise, infrared and blackbody radiation, etc., which will be explored and better understood during SuperCDMS SNOLAB.


\bigskip
\noindent 
\underline{RES-NOVA}: 
RES-NOVA is a newly proposed experiment for the detection of neutrinos from astrophysical sources~\cite{Pattavina:2020cqc}. RES-NOVA will employ an array of archaeological Pb-based cryogenic detectors sensitive to SN neutrino emission from the entire Milky Way Galaxy. Its modular design will be suited for the detection of nearby SN explosions ($<$3~kpc)~\cite{RES-NOVA:2021gqp}.

\bigskip
\noindent 
\underline{SuperNova Early Warning System (SNEWS)}: 
SNEWS is an international network of neutrino detectors in operation since 1998, with the aim of providing high-confidence alerts of nearby  core-collapse supernovae by requiring a temporal coincidence of neutrino bursts between detectors \cite{Antonioli:2004zb}. Participating facilities include not only large water-based detectors such as Super-Kamiokande and IceCube, but also scintillator-based detectors such as KamLAND and lead-based neutrino detection such as HALO. Alerts are automated, rapid, and do not require human intervention. SNEWS~2.0 is an ongoing upgrade of the SNEWS system for the multi-messenger astronomy era \cite{SNEWS:2020tbu}: false alarms are acceptable, low probability events should be reported, and SNEWS will be
one of many multi-messenger alert systems. Among the major upgrades, SNEWS~2.0 will provide pointing information obtained via triangulation, provide a pre-collapse alert obtained via pre-supernova neutrino detection, and add large-volume dark matter
detectors to the suite of growing detectors connected to the network. Among other benefits, dark matter detectors observe flavor-blind CE$\nu$NS events, which can help disentangle supernova neutrino oscillation effects when combined with existing inelastic interaction channels.

\bigskip
\noindent 
\underline{Paleo detectors}: Paleo-detectors use the nuclear damage tracks recorded in natural minerals over geological time-scales to detect weakly-interacting particles over exposures much larger than what is feasible in conventional terrestrial detectors \cite{Baum:2018tfw,Drukier:2018pdy}. Unlike conventional experiments which measure nuclear recoils in real time, paleo-detectors measure the number of events integrated over the age of the mineral, reaching up to a billion years for minerals routinely found on Earth. The sources of the keV-scale nuclear recoils which can be recovered as tracks is rich, including atmospheric neutrinos \cite{Jordan:2020gxx}, solar neutrinos \cite{Tapia-Arellano:2021cml}, supernova neutrinos \cite{Baum:2019fqm}, as well as dark matter \cite{Baum:2021jak,Baum:2021chx}. Low-energy neutrino and dark matter tracks are initiated dominantly via CE$\nu$NS (quasi-elastic charged-current interactions are more applicable for high-energy neutrinos). 

\bigskip
\noindent 
\underline{CYGNUS}: 
CYGNUS is a proposed modular and multi-site network of large-scale gas time projection chambers~\cite{Vahsen:2020pzb}. The primary goal of the CYGNUS experiment is to perform a direction-dependent search for dark matter, which has been shown to be one of the only ways to convincingly prove the galactic origin of a detected signal~\cite{Mayet:2016zxu,Vahsen:2021gnb}. Directionality is also the best means of circumventing the neutrino fog~\cite{Grothaus:2014hja,OHare:2015utx}, but requires that good performance can be achieved at the sub-10-keV nuclear recoil energies where the majority of CE$\nu$NS events coming from solar neutrinos would lie. 

The CYGNUS collaboration was formed from several smaller groups who have successfully run smaller-scale gas TPCs, similar to the modules that would eventually make up a larger coordinated experiment. An initial feasibility study was performed in Ref.~\cite{Vahsen:2020pzb}, which outlined some of the major challenges involved in building a cost-effective experiment with the sensitivity to both dark matter and neutrinos. The balancing act is in optimizing the detector have both a large enough target mass to \emph{reach} the neutrino fog---preferring large readout planes and high gas densities---while also maintaining good-enough directionality at low energies to probe \emph{through} the neutrino fog---which tends to prefer the opposite. Maximizing directional performance is crucial not just to enable good discrimination between dark matter and neutrino recoils, but also for discriminating nuclear recoil tracks from electron tracks, which ultimately sets the threshold. A nuclear recoil threshold of 8 keV has already been shown to be feasible in the 755:5 He:SF$_6$ atmospheric pressure gas mixture suggested by Ref.~\cite{Vahsen:2020pzb}. This could be lowered to 3--5 keV with further gas/readout optimisation, and the development of specialized track-fitting techniques to improve particle identification at low energies. This would enable a CYGNUS-1000 m$^3$ detector to see between 30--50 CE$\nu$NS events over a few years. CYGNUS-1000 would also be able to detect SN neutrinos from explosions within $\sim$3 kpc. Thanks to the directional sensitivity, CYGNUS may be able to point back to those SN as well~\cite{Vahsen:2020pzb}, something that would be impossible in other dark matter detectors. 

The current plans are for the individual groups involved in CYGNUS to move towards 1 m$^3$-scale prototypes to explore several R\&D directions, both in terms of gas optimization, but also to test the highly segmented charge readouts that will be required. If a high-definition prototype can demonstrate good performance, this will pave the way for a full 10 m$^3$ CYGNUS module on a 5--10 year timescale. A modular configuration is a key feature of the envisioned design, not just to avoid issues of available space underground, but more crucially because of the fact that its dimension along the drift direction has to be kept small to limit diffusion. If successful, a 10 m$^3$ TPC could in principle be placed close to a neutrino source for a dedicated direction-sensitive \cevns experiment, but in the next few years the primary plans for the collaboration will be to determine how best to scale this up to a much larger experiment. Eventually a 1000 m$^3$-scale experiment would consist of multiple modules within some common shielding, and distributed across the various sites involved in the CYGNUS project including LNGS, SUPL, Kamioka and Boulby. 

\bigskip
\noindent 
\underline{DRIFT}: 
The goal of the Directional Recoil Identification From Tracks (DRIFT) collaboration was the detection of a directional signal from Weakly Interacting Massive Particle (WIMP), halo, dark matter~\cite{PhysRevD.61.101301}. In order to accomplish this goal a unique, low-pressure, Negative Ion Time Projection Chamber (NITPC) technology was developed. The negative-ion drift allowed DRIFT NITPCs to have the lowest energy threshold and best inherent directional sensitivity of any limit-setting, directional dark matter detector. In addition, all of DRIFT's recent limits have been background-free. As a consequence, DRIFT's sensitivity to dark matter is almost 1,000 times better than other directional WIMP detectors~\cite{battat2017low}.

With its unique directional and background rejection capabilities, the DRIFT NITPC technology is ideally suited to search for nuclear recoils in beam dump experiments. Previous work involved searching for light dark matter recoils behind an electron beam-dump at JLab. Preliminary work, including a test run at SLAC, suggests that a Beam Dump experiment using a DRIFT detector, BDX-DRIFT, would have sensitivity rivaling the best limits on light dark matter and provide an unequivocal directional signature in the event of discovery~\cite{PhysRevD.99.061301}. Placing a BDX-DRIFT detector behind a proton beam dump, such as in the DUNE Near Detector Complex, is perhaps even more interesting.

The Near Detector Complex is $\sim$~100 m underground. The beam timing structure at the NuMI beam is such that backgrounds are expected to be reduced to negligible levels. Proton beam-dumps produce a plethora of neutrinos, particularly the LBNF-Dune beam, which is optimized for neutrino production. Thus, in addition to traditional beam-dump searches for light dark matter we can also search for beyond the standard model (BSM) neutrino interactions. We estimate that a 1 m$^3$ $\nu$BDX-DRIFT detector run for one year in the DUNE Near Detector Complex would detect several coherent neutrino-nucleus elastic scatters, potentially confirming recent Coherent Elastic Neutrino-Nucleus Scattering (CE$\nu$NS) detection results, but with minimal background. Off-axis and directional sensitivity will provide $\nu$BDX-DRIFT signatures to search for physics even in the presence of a neutrino background and opening up a new window to search for BSM physics.

In the near term a 1 m$^3$ $\nu$BDX-DRIFT detector is available to be deployed in the NuMI beam at Fermilab on a year or two timescale. Knowledge gained from those runs will inform proposed a proposed experiment in DUNE in the future~\citep{AristizabalSierra:2021uob}.

\bigskip
\noindent 
\underline{BULLKID} is a R\&D project on cryogenic detectors for CE$\nu$NS and light Dark Matter~\cite{colantoni:2020}. By exploiting the high multiplexing levels of kinetic inductance detectors,  goal of BULLKID is to create a monolithic and highly-segmented array of silicon targets with energy threshold on nuclear recoils around 100~eV  and total mass of 30-60~g. In future experiments several arrays would be produced and stacked to obtain target masses exceeding 1~kg.

\section{Connection with U.S. neutrino and dark matter programs}
\label{sec:connections} 
\par This whitepaper has focused on the detection of~\cevns, and the physics that may be extracted from these detections, using both terrestrial and astrophysical sources. The emphasis is on the multi-faceted experimental effort on-going around the world to expand upon the recent COHERENT measurements and to study \cevns using a wide variety of neutrino sources and detector technologies. Given the broad scientific applications of \cevns, and its complementarity to many different aspects neutrino physics, it will be an important component of the neutrino physics program in the coming decade.

\par More generally,~\cevns~experiments have broad overlap with the larger neutrino and dark matter physics programs. It has long been realized that detectors searching for WIMP dark matter are ideal~\cevns~experiments via the detection of astrophysical neutrinos. There is also overlap between~\cevns~experiments and searches for dark matter in the sub-GeV mass regime. In this context, the~\cevns represents a background to a possible dark matter signal, and therefore a precision measurement of the cross section is required.   

\par We are now just in the beginning of a very exciting time in~\cevns research. The potential to reach for new physics, or to approach questions from different angles and will different probes, will be a driving force in this field for many years to come. 

\bibliographystyle{apsrev}

\end{document}